\documentclass[12pt]{article}
\usepackage{cite}
\usepackage{graphicx}
\usepackage{amssymb}
\usepackage{amsmath}
\usepackage{amsfonts}
\usepackage{dsfont}
\usepackage{mathtools}
\usepackage{slashed}
\usepackage{rotating}
\usepackage{bbold,amsfonts}

\usepackage[utf8]{inputenc}
\usepackage{bm}
\usepackage{xcolor}
\usepackage{float}
\usepackage{braket}

\usepackage[height=8.8in,width=6.45in]{geometry}
\usepackage[font=small,labelfont=bf]{caption}
\usepackage[hidelinks]{hyperref}
\numberwithin{equation}{section}
\newcommand{\beq}{\begin{equation}}
\newcommand{\eeq}{\end{equation}}

\newcommand{\pa}{\partial}

\makeatletter
\newcommand*{\letterdef@}{}
\newcommand*{\letterdef}[3]{%
	\def\letterdef@##1{\expandafter\newcommand\csname #1\endcsname{#2{##1}}}%
	\@tfor\@tempa :=#3\do{\expandafter\letterdef@\expandafter{\@tempa}}}
\makeatother
\letterdef{c#1} {\mathcal}{ABCDEFGHIJKLMNOPQRSTUVWXYZ} % \cX = \mathcal{X}
\letterdef{rm#1}{\mathrm} {dDeimM} % \rmX = \mathrm{X} for X in {dDmM}

\newcommand{\be}{\begin{equation}}
\newcommand{\ee}{\end{equation}}
\def\bea{\begin{eqnarray}}
\def\eea{\end{eqnarray}}

\begin{document}
\begin{titlepage}
\vbox{
    \halign{#\hfil         \cr
           } % end of \halign
      }  % end of \vbox
\vspace*{15mm}
\begin{center}
{\Large \bf 
On BPS Strings in ${\mathcal N}=4$ Yang-Mills Theory
}

\vspace*{15mm}

{\large Sujay~K.~Ashok, Varun Gupta and Nemani V. Suryanarayana}
\vspace*{8mm}

Institute of Mathematical Sciences, \\
          Homi Bhabha National Institute (HBNI),\\
		 IV Cross Road, C.I.T. Campus, \\
	 Taramani, Chennai, India 600113
	 
\vskip 0.8cm
	{\small
		E-mail:
		\texttt{sashok,varungupta,nemani@imsc.res.in}
	}
\vspace*{0.8cm}
\end{center}

\begin{abstract}

We study singular time-dependent $\frac{1}{8}$-BPS configurations in the abelian sector of ${{\mathcal N}= 4}$ supersymmetric Yang-Mills theory that represent BPS string-like defects in ${{\mathbb R}\times S^3}$ spacetime. Such BPS strings can be described as intersections of the zeros of holomorphic functions in two complex variables with a 3-sphere. We argue that these BPS strings map to $\frac{1}{8}$-BPS surface operators under the state-operator correspondence of the CFT. We show that the string defects are holographically dual to noncompact probe D3-branes in global $AdS_5\times S^5$ that share supersymmetries with a class of dual-giant gravitons. For simple configurations, we demonstrate how to define a good variational problem and propose a regularization scheme that leads to finite energy and global charges on both sides of the holographic correspondence.

\end{abstract}
\vskip 1cm
	{
%		Keywords: {Defects, }
	}
\end{titlepage}

\setcounter{tocdepth}{2}
\tableofcontents
\vspace{1cm}
\begingroup
\allowdisplaybreaks

\section{Introduction}

Surface operators of Gukov and Witten \cite{Gukov:2006jk, Gukov:2008sn} are defined via surface defects in Euclidean gauge theories. These are solutions to the generalized Bogomolny equations \cite{Kapustin:2006pk} that are singular along two-dimensional subspaces.
Just as line operators provide valuable non-perturbative information about the phase structure of gauge theories \cite{tHooft:1977nqb, tHooft:1981bkw}, surface operators are expected to be useful in capturing novel non-perturbative physics. For instance, it has been shown in \cite{Gukov:2013zka} that surface operators can be used as order parameters for topological phases that could not be distinguished by the usual line operators. These and many other related results justify a more detailed study of surface defects.  

In this work we adopt a Hamiltonian perspective and study (at a classical level) two-dimensional defects in the maximally supersymmetric ${\mathcal N}=4$ Yang-Mills theory on $\mathbb{R}\times S^3$ spacetime as classical singular solutions  that preserve some supersymmetry. Henceforth we refer to such solutions as BPS strings.
We will focus our attention on a particularly interesting class of BPS strings that preserve four supersymmetries. Our goal in this work on the field theory side is twofold. Firstly, to find a general characterization of these BPS strings by describing the equations defining their worldvolume in a compact way. Secondly, to show that these BPS strings are solutions to the same variational problem as other non-singular supersymmetric solutions in the theory and to calculate their (regularized) energies and charges. 

As a first step we choose a suitable set of supersymmetries that we would like our solutions to preserve. For this we adopt a bottom-up approach by proposing simple classical half-BPS string solutions and determine their supersymmetries as projection conditions on the conformal Killing spinors of $\mathbb{R}\times S^3$. These half-BPS strings are static configurations, with topology $\mathbb{R}\times S^1$. By using the state operator correspondence, we show that these BPS strings are the states that correspond to the half-BPS Gukov-Witten surface operators in $\mathbb{R}^4$. By using global symmetries, we find more such half-BPS string solutions and observe that all these defects have two supersymmetries in common. The common supersymmetries can be used to derive a set of non-abelian BPS equations whose solutions are at least $\frac{1}{16}$-BPS. It turns out that these BPS equations coincide with those of \cite{Grant:2008sk, Yokoyama:2014qwa, SARPNVS} obtained in the study of the gauge theory duals of giant gravitons and dual-giant gravitons in $AdS_5\times S^5$ \cite{McGreevy:2000cw, Grisaru:2000zn, Hashimoto:2000zp}. 
In fact, we find that the time dependent non-singular classical configurations dual to half-BPS dual-giants share a common set of four supersymmetries with the half-BPS strings supported by one complex scalar field. We then go on to derive the general non-abelian $\frac{1}{8}$-BPS equations that bosonic configurations have to satisfy in order to preserve these four supersymmetries.\footnote{We mention that a similar exercise has been carried out recently in \cite{Wang:2020seq} by topologically twisting the ${\mathcal N}=4$ Yang-Mills theory. The cohomology of the chosen $Q_{BRST}$ operator includes surface defects, line defects and local operators (see also \cite{Gutperle:2019dqf, Gutperle:2020gez, Komatsu:2020sup} for recent work related to surface defects).} 

The resulting $\frac{1}{8}$-BPS equations are what we focus on in this work and for the most part we restrict our analysis to abelian solutions in the scalar sector. One of our main results is a simple characterisation of the world-volumes of the time-dependent $\frac{1}{8}$-BPS strings, which we shall also refer to as wobbling strings. We show that at any given time the spatial configuration of the wobbling string is obtained as the intersection of the zeros of a holomorphic function $F(z_1, z_2) =0$ with the 3-sphere defined by $|z_1|^2 + |z_2|^2 = 1$ with its time evolution obtained by $(z_1, z_2) \rightarrow (z_1 \, e^{-i\tau}, z_2 \, e^{-i \tau})$. This is the general characterization we were after. Then we show that these BPS strings can be obtained as solutions to a well-defined variational problem, by adding particular boundary terms at the location of the string. We then focus on a sub-class of solutions that correspond to functions $F(z_1,z_2)$ that are of the monomial type. Naively the energy and global charges of these singular configurations appear to diverge if we just use the ${\mathcal N}=4$ gauge theory Lagrangian. However we show that by adding further boundary terms (that do not affect the variational problem), the energy and other global charges of these wobbling string solutions can be made finite. 

We then turn to the holographic approach to the study of these string solutions, by studying probe D3-branes in $AdS_5\times S^5$.  
For half-BPS defects in ${\cal N}=4$ SYM in $\mathbb{R}^4$, the holographic duals have been obtained in \cite{Gomis:2007fi, Drukker:2008wr} as both bubbling geometries as well as probe D3-branes. This has been generalized to defects that preserve fewer number of supersymmetries in \cite{Koh:2008kt}. We consider various classes of $\frac{1}{2}$-BPS probe D3-branes in global $AdS_5\times S^5$: the equations that define the worldvolume of these probes are largely inspired by the profiles of the scalar fields of the half-BPS strings in the boundary gauge theory on ${\mathbb R} \times S^3$. These are noncompact probe branes that end on the boundary in $\mathbb{R}\times S^1$. The intersection of the D3-brane probe with the boundary is essentially the half-BPS string of the ${\mathcal N}=4$ theory.  

We then mirror the analysis of the boundary theory and perform a $\kappa$-symmetry analysis to find the projections on the bulk Killing spinor for the various $\frac{1}{2}$-BPS probes. Remarkably, we find that the set of supersymmetries common to all these static defects coincides precisely with those preserved by the most general giant and dual-giant configurations in $AdS_5\times S^5$ derived in \cite{Kim:2006he, Ashok:2008fa}. The worldvolumes of such probe branes are known to be described in terms of zeros of holomorphic functions. For the holographic duals of the $\frac{1}{8}$-BPS wobbling strings, we show that near the boundary of $AdS_5$, the zero locus of the holomorphic function coincides with the location of the BPS string of the boundary theory and proceed to derive the singular boundary scalar field profiles from the D3-brane solutions. We thereby recover the general characterization of the wobbling strings from a probe analysis in the bulk dual. 

Finally we restrict attention to the D3 branes dual to the monomial type BPS strings of the CFT. By adding an appropriate set of boundary terms we define a variational problem that admits all such brane configurations as allowed solutions. We then carry out the holographic renormalization of energies and charges in an expansion around the large energy limit of the probe brane. We are able to match the expected boundary results in the leading approximation and we go on to obtain the first order correction to the Yang-Mills results. The holographic renormalization we carry out in the bulk closely resembles the analogous calculation in the boundary theory and provides a justification for the regularization we carry out in the boundary theory. 

This paper is organized as follows. In Section \ref{defectsRS3} we study BPS strings in ${\mathcal N}=4$ Yang-Mills theory on $\mathbb{R}\times S^3$. We perform a detailed supersymmetry analysis of the various classical configurations and the final result is a derivation of a particular class of $\frac{1}{8}$-BPS equations. 
In Section \ref{giantdefects} we characterize the abelian solutions of these equations that correspond to wobbling strings and show that they are all solutions to the same variational problem. For particularly simple monomial type solutions, we compute the renormalized energy and global charges. 
In Section \ref{D3probes} we study probe D3-branes in global $AdS_5\times S^5$ and exhibit the particular probes that are dual to the wobbling BPS strings. 
In Section \ref{wobblers} we focus on the holographic duals of the monomial solutions and compute the energy using holographic renormalization. 
We conclude in Section \ref{SandD} with a summary of our main results and a discussion of the possible implications of our results. Some technical details are collected in the appendices, along with a brief discussion of ``pure glue" defects. 

\section{\texorpdfstring{${\mathcal N}=4$}{N4} Gauge Theory on \texorpdfstring{$\mathbb{R}\times S^3$}{RS3} }
\label{defectsRS3}

We begin with the action of ${\mathcal N}=4$ Yang-Mills theory on $\mathbb{R}\times S^3$:
\begin{multline}
S = \frac{1}{g_{YM}^2}\int d^4x \,  \sqrt{-g} ~ \text{Tr} \bigg( -\frac{1}{4}F_{\mu\nu}^2 -\frac{1}{2} (D_{\mu}X_m)^2+\frac{1}{4}[X_m,X_n]^2  -\frac{1}{2}X_m^2 \\
-\frac{i}{2} \bar\lambda \Gamma^{\mu}D_{\mu}\lambda - \frac{1}{2}\bar\lambda\Gamma^m[X_m,\lambda] \bigg)
\end{multline}
Here $A_{\mu}$ is the gauge field, the $X_m$ for $m\in \{4,5, \ldots 9\}$ are the six scalars that transform in the vector representation of the SO$(6)$ R-symmetry group and $\lambda$ is the gaugino and is a ten dimensional Majorana-Weyl fermion. All fields transform in the adjoint representaton of the gauge group U($N$). The $D_{\mu}$ are the covariant derivatives that are both gauge and general covariant. To clarify our conventions we now discuss the geometry of the background in detail.  

\subsection{The Geometry of \texorpdfstring{$\mathbb{R}\times S^3$}{RS3}}

We choose the following metric on $\mathbb{R}\times S^3$:
\begin{equation}
ds^2 = -d\tau^2 + d\theta^2 + \cos^2 \theta \, d\phi_1^2 + \sin^2\theta \, d\phi_2^2
\end{equation}
These are the natural coordinates that arise while taking the boundary limit of the bulk metric in \eqref{gddbulk}. However we now define more convenient coordinates in terms of which one writes the three sphere as a Hopf fibration over the two sphere. We define the angles $\psi = \phi_1 + \phi_2$, $\varphi = \phi_1 - \phi_2$ and $\vartheta = 2\theta$, in terms of which the metric takes the form 
\be\label{hopf}
ds^2=-d\tau^2 + \frac{1}{4}((d\psi+\cos\vartheta \, d\varphi)^2 + d\vartheta^2 + \sin^2\vartheta \, d\varphi^2)\,.
\ee
Here, $\psi$ is the Hopf-fibre coordinate and $(\vartheta, \varphi)$ specifying the directions of the two sphere. We will use both these two coordinate systems interchangeably. 

A natural choice of one-forms on this manifold is given by $e^0 = d\tau$ and the right-invariant one-forms on the three sphere:
\be
\begin{aligned}
\label{hopfvierbein}
e^1 &= \frac{1}{2}(-\sin\psi \, d\vartheta + \cos\psi \sin\vartheta \, d\varphi)\,, ~\,
e^2 = \frac{1}{2}(\cos\psi \, d\vartheta + \sin\psi \sin\vartheta \, d\varphi)~,\\ 
\text{and}\qquad e^3 &= \frac{1}{2} (\cos\vartheta \, d\varphi + d\psi)\,.
\end{aligned}
\ee
In terms of these, the metric can be simply written as
$
g_{\mu\nu}= \eta_{ab}e^a_{\mu}e^{b}_{\nu} \,,
$
where $\eta = \text{diag}(-1,1,1,1)$. The vierbein satisfy the relations 
\begin{equation}
de^0 = 0, ~~~ de^a = {\epsilon^a}_{bc} e^b \wedge e^c~,\quad \text{for}\quad {a,b,c} \in \{1,2,3\}~.
\end{equation}
This allows us to read-out the non-zero components of the spin connection ${\omega^a}_b = {{\omega_\mu}^a}_b dx^\mu$ to be
\begin{equation}
{\omega^a}_b = {\epsilon^a}_{bc} e^c\quad \text{for} \quad{a,b,c} \in \{1,2,3\}~.
\end{equation}

In what follows we will use the isomorphism between SO$(6)$ and SU$(4)$ and rewrite the action in the SU$(4)$ covariant notation. We will follow the conventions of \cite{Ishiki:2006rt} and define the complex matrix $X$, which will be parametrized by the six real scalars $X^m$:
\be
\label{XABdefn}
X = \begin{pmatrix}
0  & Z_3^{\dagger} & -Z_2^{\dagger} & Z_1\\
-Z_3^{\dagger}&0 & Z_1^{\dagger} & Z_2\\
Z_2^{\dagger} & -Z_1^{\dagger} & 0  & Z_3 \\
-Z_1 & -Z_2 & -Z_3 & 0
\end{pmatrix} \,.
\ee
The entries of this matrix will be denoted $X_{AB}$, where the indices $A, B \in \{1,2,3,4\}$ are in the fundamental representaion of SU$(4)$, the R-symmetry group.  Here $X_{AB} = -X_{BA}$ and it follows that it transforms as the ${\bf 6}$ of the SU$(4)$ group. This is related to the matrix $X^{AB}$ with raised indices by the relation
\be
X^{AB} = \frac{1}{2}\epsilon^{ABCD} X_{CD} \,,
\ee
where the $\epsilon$ is the completely anti-symmetric tensor. 
The $Z_i$ that appear as the entries of the matrix $X$ are the following complex combinations of the scalars:
\be
Z_1 =  \frac{1}{2}(X_4+ i X_5) \quad Z_2 = \frac{1}{2}(X_6+ i X_7) \quad Z_3 =  \frac{1}{2}(X_8+ i X_9) \,. 
\ee
To rewrite the action and supersymmetry variations in an SU$(4)$ invariant form, we need to decompose the ten dimensional Majorana-Weyl spinor in an appropriate manner. We refer the reader to \cite{Ishiki:2006rt} for details and we merely present the results. The ten dimensional spinor is decomposed as follows:  
\be
\label{10dMW}
\epsilon = \begin{pmatrix}
\epsilon_+^A\\
\epsilon_{-A}
\end{pmatrix}\,,
\ee
where $\epsilon_{-A}$ is the charge conjugate of $\epsilon_+^A$. The $\pm$ subscript indicates the four-dimensional chirality of the spinors, $\gamma_5 \epsilon_{\pm} = \pm \epsilon_{\pm}$. 

The action can now be written in SU$(4)$ invariant form. In addition, our analysis becomes much easier if we express all vector quantities in terms of the tangent space indices, using the viebein in \eqref{hopfvierbein}:
\be
A_{\mu} = e_{\mu}^a A_a~, \quad \gamma^{\mu} = e^{\mu}_a \Gamma^a~,\quad D_{\mu} = e^{a}_{\mu}D_{a}~,\text{etc.}
\ee
Then, the action takes the following form:
\be
\label{action2}
\begin{aligned}
S = \frac{1}{g_{YM}^2}\int d^4x \, e \, \text{Tr} \bigg(&-\frac{1}{4}F_{ab}^2 -\frac{1}{2} D_{a}X_{AB} D^{a}X^{AB}+\frac{1}{4}[X_{AB},X_{CD}][X^{AB},X^{CD}]  \cr 
&-\frac{1}{2}X_{AB}X^{AB}-i\bar\lambda_{+A}\Gamma^{a}D_{a}\lambda_+^A - \bar\lambda_{+A}[X^{AB},\lambda_{-B}] - \bar\lambda_-^A[X_{AB},\lambda_+^B] \bigg)
\,.
\end{aligned}
\ee
We note that the covariant derivative $D_a$ is both gauge and local Lorentz covariant. Further the local Lorentz basis is simply dual to the vierbein in \eqref{hopfvierbein}. Explicitly, we have the following expressions:
\begin{align}
\begin{split}
E_0:= e_0^\mu \partial_\mu &= \partial_\tau, \\
E_1:= e_1^\mu \partial_\mu &= 2 \, [-\sin\psi \, \partial_\vartheta + \cos\psi \, (\csc\vartheta \partial_\phi - \cot\vartheta \, \partial_\psi)]\\
E_2:= e_2^\mu \partial_\mu &= 2 \, [\cos\psi \, \partial_\vartheta + \sin\psi \, (\csc\vartheta \partial_\phi - \cot\vartheta \, \partial_\psi)]\\
E_3:=e_3^\mu \partial_\mu &= 2\, \partial_\psi 
\end{split}
\label{hopfvectorfields}
\end{align}
These vector fields satisfy the following commutation relations:
\be
[E_0, E_a] = 0, ~~ [E_a, E_b] = -2 {\epsilon_{ab}}^c \, E_c ~~~ {\rm for}\quad a,b,c \in \{1,2,3\}~.
\ee

\subsection{Supersymmetry Variations and Conformal Killing Spinors}

The action (\ref{action2}) is invariant under the following supersymmetry variations and we shall present these in the SU$(4)$ notation:
\begin{align}
\begin{split}
\delta A_{a} &= i(\bar\lambda_{+A}\Gamma_{a}\epsilon_+^A - \bar\epsilon_{+A}\Gamma_{a}\lambda_+^A)\\
\delta X^{AB} &= i(-\bar\epsilon_-^A\lambda_+^B + \bar\epsilon^B_-\lambda_+^A + \epsilon^{ABCD} \bar\lambda_{+C}\epsilon_{-D} ) \\
\delta\lambda_+^A & = \frac{1}{2}F_{ab}\Gamma^{ab}\epsilon_+^A + 2 D_{a}X^{AB}\Gamma^{a}\epsilon_{-B} + X^{AB}\Gamma^{a}\nabla_{a}\epsilon_{-B} + 2i[X^{AC},X_{CB}]\epsilon_+^B\\ 
\delta\lambda_{-A} &=\frac{1}{2}F_{ab}\Gamma^{ab}\epsilon_{-A} + 2 D_{a}X_{AB}\Gamma^{a}\epsilon_{+}^{B} + X_{AB}\Gamma^{a}\nabla_{a}\epsilon_+^{B} + 2i[X_{AC},X^{CB}]\epsilon_{-B} \,.
\end{split}
\end{align}
The $\epsilon_{\pm, A}$ are conformal Killing spinors on $\mathbb{R}\times S^3$. The subscript $\pm$ refers to the four dimensional chirality and the SU$(4)$ index $A$ indicates that there are four such spinors of each chirality. Each of the epsilons account for four independent real parameters and thus, the ${\mathcal N}=4$ gauge theory has 32 supersymmetries which can equivalently be encoded in the ten dimensional Majorana-Weyl spinor shown in \eqref{10dMW}. 

The conformal Killing spinor (CKS) of negative 4d chirality satisfies the following equation \cite{Ishiki:2006rt}:
 \begin{align}
 \label{KSeqn1}
 \nabla_{a} \epsilon^{(\pm)}_{-A} = \pm \frac i2 \Gamma_{a} \Gamma^0 \epsilon^{(\pm)}_{-A} \,,
 \end{align}
 where $\epsilon^{(\pm)}_{-A}$ have the same chirality on account of the projection $i\, \Gamma_{0123} \, \epsilon^{(\pm)}_{-A}=  \, \epsilon^{(\pm)}_{-A} $.
Using our choice of vierbein, one can solve for the CKS equation and we find the following solutions:
 \begin{subequations}
 \begin{align}
 \label{KSR3XSone}
 \epsilon^{(-)}_{-A} &= e^{- \frac{i\,\tau}{2}} \eta^{(-)}_A \,,\\
 \epsilon^{(+)}_{-A} &= N \cdot \eta^{(+)}_A = e^{ \frac{i\,\tau}{2}} \, e^{- \frac{\Gamma_{12}}2 \psi} \, e^{ - \frac{\Gamma_{31}}2 \vartheta} \, e^{- \frac{\Gamma_{12}}2 \phi}\eta^{(+)}_A \,,
 \label{KSR3XStwo}
 \end{align}
 \end{subequations}
where we have defined the matrix $N$ and the $\eta^{(\pm)}_A$ are the constant spinors that satisfy the 4d chirality constraint $ i\, \Gamma_{0123} \, \eta^{(\pm)}_A=  \, \eta^{(\pm)}_A $. 
 
\subsection{\texorpdfstring{$\frac{1}{2}$}{12}-BPS Configurations} 

We now exhibit different bosonic solutions to the equations of motion and show that they all preserve half of the supersymmetries. We do this by using the explicit solutions for the scalar fields and deriving the projections on the constant spinors that follow from setting the supersymmetry variations of gauginos to zero. Once we get these projections, then, we combine them in interesting ways to find non-abelian BPS equations that are preserved by the intersection of these projections.

\subsubsection{A First Class of Stringy Defects} 
 
We now propose the following non-trivial (singular) classical configuration:
\be
\begin{aligned}
\label{class1def}
Z_1 &= \frac{c_1}{\cos\theta \, e^{i\phi_1}}  = c_1\, \sec\frac{\vartheta}{2} \, e^{-\frac{i}{2}(\varphi+\psi)} \\
F_{ab} &=Z_2= Z_3=0~,
\end{aligned}
\ee
where $c_1$ is a Cartan generator of the gauge group, and we have expressed the solution in both sets of coordinates. 
One can check that the proposed solution satisfies the equations of motion. Further, for this abelian solution, one can check that it satisfies the following equations:
\be
\label{DZ1eqns}
D_0 Z_1 = 0\,, \quad (D_3 + i)Z_1 = 0\,, \quad (D_1+ i D_2)Z_1 = 0 \,.
\ee
Since the gauge field is set to zero, the $D_a$ in the Lorentz basis simply coincide with the vector fields $E_a$ given in \eqref{hopfvectorfields}. 
These in turn will aid us in verifying that this is a $\frac{1}{2}$-BPS solution. Since it is a purely bosonic and abelian solution, we only have to check that the following gaugino variation is zero on the solution:
\be
2 D_{a}X^{AB}\Gamma^{a}\epsilon_{-B} + X^{AB}\Gamma^{a}\nabla_{a}\epsilon_{-B}  = 0 ~.
\ee
For the first solution in  \eqref{KSR3XSone}, we find the following projections on the constant spinor:
\be
\begin{aligned}
\label{Donem}
(1+\Gamma^{03}) \, \eta^{(-)}_A &= 0 \quad\text{for}\quad A=1, 4~\\
(1-\Gamma^{03}) \, \eta_A^{(-)} &= 0 \quad\text{for}\quad A= 2, 3~.
\end{aligned}
\ee
For the other conformal Killing spinor, the analysis is a little more involved as one has to move the $\Gamma$-matrices through the coordinate dependent matrix, which we denoted $N$ in \eqref{KSR3XStwo}. 
The following identities prove to be useful:
\be
 \begin{aligned}
    \Gamma_1\, N 
    %&= e^{\Gamma_{12} \frac{\psi}{2}} e^{-\Gamma_{13} \frac{\theta}{2}} e^{\Gamma_{12} \frac{\varphi}{2}} \, \Gamma_1 \\
   &= N \big(\cos \psi \sin \vartheta \, \Gamma_3 - \left(  \sin \psi \cos \varphi + \cos \psi \cos \vartheta \sin \varphi  \right) \Gamma_2 \\
&\hspace{6cm}   -  \left(  \sin \psi \sin \varphi - \cos \psi \cos \vartheta \cos \varphi  \right) \Gamma_1 \big) ~,\\
    \Gamma_2\, N
    %&= e^{\Gamma_{12} \frac{\psi}{2}} e^{\Gamma_{13} \frac{\vartheta}{2}} e^{\Gamma_{12} \frac{\varphi}{2}} \, \Gamma_2 \\
    &= N \big( \sin \psi \sin \vartheta \, \Gamma_3  + \left( \cos \psi \cos \varphi - \sin \psi \cos \vartheta \sin \varphi \right) \Gamma_2 \\
&    \hspace{6cm}+\left( \cos \psi \sin \varphi + \sin \psi \cos \vartheta \cos \varphi  \right) \Gamma_1 \big)~, \\
     \Gamma_3\, N
     %&= e^{-\Gamma_{12} \frac{\psi}{2}} e^{-\Gamma_{13} \frac{\vartheta}{2}} e^{-\Gamma_{12} \frac{\varphi}{2}} \, \Gamma_3 \,\, 
     &= N \big( - \sin \vartheta \cos \varphi \, \Gamma_1 + \sin \vartheta \sin \varphi \, \Gamma_2 + \cos \vartheta \, \Gamma_3 \big) ~.
     \end{aligned}
\ee
Substituting these into the supersymmetry variations and setting them to zero, we find the projections: 
\be
\begin{aligned}
(1-\Gamma^{03}) \, \eta^{(+)}_A &= 0 \quad\text{for}\quad A=1, 4~\\
(1+\Gamma^{03}) \, \eta_A^{(+)} &= 0 \quad\text{for}\quad A= 2, 3~.
\end{aligned}
\label{Donep}
\ee
We have thereby shown that the proposed solution (\ref{class1def}) is half-BPS as it preserves exactly half of the supersymmetries of the gauge theory.

 \begin{figure}[!htb] 
 \center{\includegraphics[width=\textwidth] {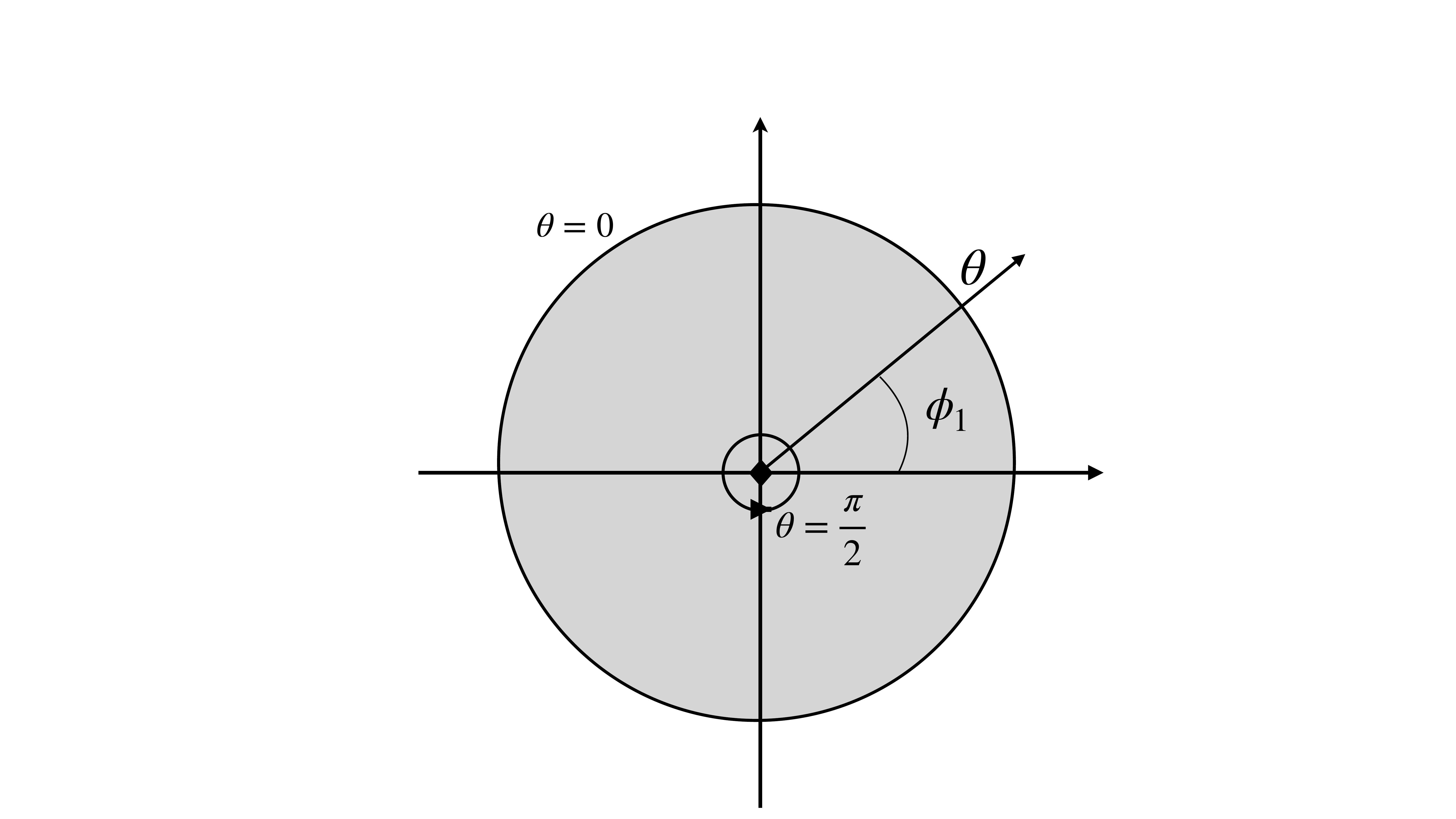}} 
 \caption{\label{fig:my-label}The topology of the space transverse to the defect is a disk. At the center of the disk we have $\theta=\frac{\pi}{2}$ and at the boundary of the disk we have $\theta=0$.}  
 \label{thetaphi1}
 \end{figure}

We interpret this solution as a monodromy defect on $\mathbb{R}\times S^3$, analogous to the Gukov-Witten defect in $\mathbb{R}^4$. The defect is extended along the $(\tau, \phi_2)$ directions while the two directions transverse to the defect are parametrized by $(\theta, \phi_1)$ coordinates. The transverse space has the topology of a disk, as shown in Figure \ref{thetaphi1}. The constant matrix $c_1$ that appears in the classical solution encodes the $(\beta, \gamma)$ parameters that appears in the Gukov-Witten solution \cite{Gukov:2006jk}. In the U$(N)$ theory, we can write down the following generalized solution for the scalar profile: 
\begin{equation}
Z_1= \begin{pmatrix}
		c_{1,1}\,\mathbb{I}_{n_1}&0&\cdots&0\\
		0&c_{1,2}\,\mathbb{I}_{n_2}&\cdots&0\\
		\vdots&\vdots&\ddots&\vdots\\
		0&0&\cdots&c_{1,M}\,\mathbb{I}_{n_{M}}
		\end{pmatrix} \frac{1}{\cos\theta \, e^{i\phi_1}}  ~.
\end{equation}
In addition, it is possible to turn on an independent parameter for the gauge field that corresponds to a non-trivial holonomy for the gauge field, with $A = \underline{\alpha}\, d\phi_1$, where $\underline{\alpha}$ is an element of the Cartan subalgebra that breaks the U$(N)$ to the subgroup U$(n_1)\times $U$(n_2) \times \ldots $U $(n_M)$. The $\underline{\alpha}$-parameters encode the monodromy of the four dimensional gauge field around the location of the stringy defect. In the rest of our discussions in both the super-Yang-Mills theory and the holographic bulk theory, we shall not turn these parameters on and focus mostly on the scalar profiles.

\subsubsection{More Defects in the First Class}

One can get two more defects in the same class by using an SU$(3)$ rotation to change the scalar $Z_1$ to one of the others, either $Z_2$ or $Z_3$. The derivation of the projection conditions follows along the same lines and we simply present the projection conditions. 

For the defect corresponding to the scalar profile $Z_2= c_2\, \sec\frac{\vartheta}{2} \, e^{-\frac{1}{2}(\varphi+\psi)}$, we find that the half-BPS projections are given by
\be
\begin{aligned}
(1\pm\Gamma^{03}) \, \eta^{(\mp)}_A &= 0 \quad\text{for}\quad A=2, 4~\\
(1\mp\Gamma^{03}) \, \eta_A^{(\mp)} &= 0 \quad\text{for}\quad A= 1, 3~.\\
\end{aligned}
\ee
We see that it simply amounts to a permutation of the $\{1,2\}$ labels of the SU$(4)$ R-symmetry index in the projection conditions obtained previously. 

Similarly, for the defect corresponding to the scalar profile $Z_3= c_3\, \sec\frac{\vartheta}{2} \, e^{-\frac{1}{2}(\varphi+\psi)}$, we find the following projections:
\be
\begin{aligned}
(1\pm\Gamma^{03}) \, \eta^{(\mp)}_A &= 0 \quad\text{for}\quad A=3, 4~\\
(1\mp\Gamma^{03}) \, \eta_A^{(\mp)} &= 0 \quad\text{for}\quad A= 1, 2~.\\
%(1-\Gamma^{03})\eta^{(+)}_A &= 0 \quad\text{for}\quad A=3, 4~\\
%(1+\Gamma^{03})\eta_A^{(+)} &= 0 \quad\text{for}\quad A= 1, 2~.
\end{aligned}
\ee

\subsubsection{A Second Class of Stringy Defects} 

Let us now consider a second class of defects, with singular profile given by
\be
\begin{aligned}
\label{Z1sintheta}
Z_1 &= \frac{d_1}{\sin\theta \, e^{i\phi_2}}  = d_1\, \csc\frac{\vartheta}{2} \, e^{-\frac{i}{2}(\psi-\varphi)} \\
F_{ab} &=Z_2= Z_3=0~. 
\end{aligned}
\ee
One can check, as before, that the profile satisfies the equation of motion and that it also satisfies the differential equations in \eqref{DZ1eqns}. Following the same procedure as before, one can check that this solution also preserves half of all the supersymmetries.  The first set of projections, on the spinor $\eta_A^{(-)}$, is identical to the projections in \eqref{Donem} for the first class of defects: 
\be
\begin{aligned}
(1+\Gamma^{03}) \, \eta^{(-)}_A &= 0 \quad\text{for}\quad A=1, 4~\\
(1-\Gamma^{03})\, \eta_A^{(-)} &= 0 \quad\text{for}\quad A= 2, 3~.
\end{aligned}
\label{Dtwom}
\ee
However, on the $\eta^{(+)}_A$, a different set of supersymmetries is preserved and we find the following projections:
\be
\begin{aligned}
(1+\Gamma^{03}) \, \eta^{(+)}_A &= 0 \quad\text{for}\quad A=1, 4~\\
(1-\Gamma^{03}) \, \eta_A^{(+)} &= 0 \quad\text{for}\quad A= 2, 3~.
\end{aligned}
\label{Dtwop}
\ee
As before, it is possible to write a more general solution for the U$(N)$ gauge theory by making the $Z_1$ a general linear combination of the Cartan generators and by also turning on a non-trivial holonomy for the gauge field. 

\subsubsection{More Defects in the Second Class}

One can obtain two more defects in the same class by performing an SU$(3)$ rotation. For the defect corresponding to the scalar profile $Z_2= c_2\, \csc\frac{\vartheta}{2} \, e^{-\frac{i}{2}(\psi-\varphi)}$, we find that the half-BPS projections are given by
\be
\begin{aligned}
(1+\Gamma^{03}) \, \eta^{(\mp)}_A &= 0 \quad\text{for}\quad A=2, 4~\\
(1-\Gamma^{03}) \, \eta_A^{(\mp)} &= 0 \quad\text{for}\quad A= 1, 3~.\\
\end{aligned}
\ee
Similarly, for the defect corresponding to the scalar profile $Z_3= c_3\, \csc\frac{\vartheta}{2} \, e^{-\frac{i}{2}(\psi-\varphi)}$, we find the following projections:
\be
\begin{aligned}
(1+\Gamma^{03}) \, \eta^{(\mp)}_A &= 0 \quad\text{for}\quad A=3, 4~\\
(1-\Gamma^{03}) \, \eta_A^{(\mp)} &= 0 \quad\text{for}\quad A= 1, 2~.\\
\end{aligned}
\ee

\subsection{Classical BPS Equations}

Now that we have derived the projections associated to each of the classical $\frac{1}{2}$-BPS configurations, what we would like to do now is to reverse the logic. We first find out the set of supersymmetries common to {\it all} these configurations. Using these in the supersymmetry variations we shall find the most general (non-abelian) BPS equations that are implied by this common set of supersymmetries. 

First of all, we find that all of the $\eta^{(+)}_A$ spinors are projected out and we have exactly two unbroken supersymmetries, that correspond to the projection condition
\be
(1+\Gamma^{03}) \,\eta^{(-)}_4 = 0~. 
\ee
Substituting this projection into the supersymmetry variation of the gaugino $\delta\lambda_+^A$, we obtain the following BPS equations:
\be
\begin{aligned}
(D_0+D_3+i)Z_j  &= 0~,\qquad
(D_1+i D_2)Z_j = 0 \quad\text{for}\quad j=1, 2, 3~.
\end{aligned}
\label{Zeqnseta4}
\ee
A similar calculation for the variation $\delta\lambda_{-A}$ leads to the BPS equations:
 \be
 \begin{aligned}
F_{12} +  2\sum_{j=1}^3 \left[Z_j, Z_j^{\dagger} \right]  = 0~, \qquad \left [Z_i, Z_j \right] &= 0~,\\
 F_{03} = 0\,, \quad F_{01}+F_{31}=0~, \quad F_{02}+F_{32} &= 0~.
 \label{Feqnseta4}
\end{aligned}
  \ee
 It is important to recall that the $D_a$ are gauge and local Lorentz covariant derivatives on the $\mathbb{R}\times S^3$ background in the frame basis. Up to a minor change in conventions, these are precisely the $\frac{1}{16}$-BPS equations obtained in \cite{Grant:2008sk}, where they were derived using the simple Bogomolny method of writing the energy of the Yang-Mills on $\mathbb{R}\times S^3$ as a sum of squares. These equations were also derived by performing a supersymmetry analysis of $\frac{1}{16}$-BPS states in \cite{Yokoyama:2014qwa, SARPNVS}. 
 
We have arrived at the {\it same} set of equations by a supersymmetry analysis of static extended string-like defects in the gauge theory, which shows that they share unbroken supersymmetries with the BPS states found in \cite{Grant:2008sk, Yokoyama:2014qwa}. 
 
 \subsubsection{The Dual of a Dual-Giant Graviton and $\frac{1}{8}$-BPS Equations} 

In order to make explicit this point about shared supersymmetries, we now introduce our last class of half-BPS classical configurations, which is very well studied and is given by
\be
Z_1 = c ~ e^{-i\tau}~, \quad\text{and}\quad Z_2=Z_3=F_{ab} = 0~.
\ee
This time-dependent classical configuration satisfies the differential constraints
\be
(D_0 + i)Z_1= 0 \qquad D_a Z_1 = 0~.
\ee
Substituting these equations into the supersymmetry variations we find that the supersymmetry generated by the following constant spinors are preserved by this classical configuration:
\be
\begin{aligned}
\eta_A^{(-)}&\quad\text{for}\quad A=1, 4\\ 
\text{and}\quad \eta_A^{(+)} &\quad\text{for}\quad A=2, 3~.
\end{aligned}
\label{DGproj}
\ee
We now look for common supersymmetries preserved by this configuration along with defects in the first and second class that  have non-trivial $Z_1$ profile. By comparing the projection conditions in \eqref{DGproj} with the supersymmetries preserved by the defects (see \eqref{Donem}, \eqref{Donep}, and \eqref{Dtwom}, \eqref{Dtwop}), it is clear that there are common unbroken supersymmetries between these classical solutions, given by the following projections:
\be
(1+\Gamma^{03})\, \eta_{A}^{(-)} = 0~\quad\text{for}\quad A=1, 4~,
\ee
with the other $\eta_A^{(-)}$, for $A=2, 3$ and all the $\eta_A^{(+)}$ set to zero. These leave 4 unbroken supercharges, as expected for a $\frac{1}{8}$-BPS configuration. Note that these include the two supersymmetries common to all defects we considered plus two additional ones. We now impose these projection conditions on the supersymmetry variations of the fermions to find the most general $\frac{1}{8}$-BPS equations. 

We have already derived the BPS conditions that follow from the $A=4$ case and now we turn to the BPS equations that follow from imposing the projection conditions on  $\eta_{1}^{(-)}$. The equations involving the field strengths are the same as those in \eqref{Feqnseta4}. However, from the gaugino variation $\delta\lambda_+^A$, 
we find the following equations :
\be
\begin{aligned}
(D_0+D_3+i)Z_1  &= 0~,\qquad (D_1+i D_2)Z_1= 0~, \\
(D_0+D_3-i)Z_j  &= 0~,\qquad (D_1-i D_2)Z_j= 0~,\quad\text{for}\quad j=2, 3~.
\end{aligned}
\ee
The $\frac{1}{8}$-BPS equations are obtained by imposing these along with the equations in \eqref{Zeqnseta4}. As a result, we find immediately that two of the scalars are set to zero:
\be
Z_2= Z_3= 0 ~.
\ee
Thus, only those classical configurations are allowed, for which a single scalar is turned on. The remaining equations simplify and we obtain our final result for the $\frac{1}{8}$-BPS equations:
\be
\begin{aligned}
\label{finalBPS}
(D_0+D_3+i) \, Z_1  &= 0~,\quad (D_1+i D_2) \, Z_1= 0~,\quad F_{12} +  2\left[Z_1, Z_1^{\dagger} \right]  = 0~,\\
 F_{03} &= 0\, \qquad\quad F_{01}+F_{31}=0~, \qquad F_{02}+F_{32} = 0~.
\end{aligned}
\ee
For the rest of this work, we will focus on these equations and their general abelian solutions. 

\subsection{Equations of Motion and Bianchi Identities}
\label{EoMBI}

Given the half-BPS equations, it turns out that the equations of motion and the Bianchi identities are automatically satisfied. However it turns out that these lead to additional differential constraints on the gauge field if we only impose the $\frac{1}{8}$-BPS equations. Although these constraints have been discussed in \cite{Yokoyama:2014qwa, SARPNVS}, we shall find it useful to rederive them in a frame basis.  

Let us first of all begin with the scalar equation of motion. In the $\frac{1}{8}$-BPS sector of interest, in which only a single scalar is turned on, which we shall henceforth denote by $Z$, the equation of motion for the scalar field is given by
\begin{align}
D_a D^a Z + 2[Z, [Z, Z^{\dagger}]] - Z &= 0~.
\end{align}
The kinetic term can be rewritten as follows:
\be
D_a D^a Z = -D_0^2Z + D_3^2Z+(D_1-iD_2)(D_1+iD_2)Z - i [D_1, D_2]Z~.
\ee
The last term can be written as 
\be
[D_1, D_2]Z = -i[F_{12},Z]~. 
\ee
The third term is quite tricky to handle due to the non-trivial spin connection. We introduce the gauge covariant derivative ${\cal D}_a$ to be
\be
{\cal D}_\mu(\cdot) = \partial_\mu(\cdot)  - i\, \big[A_\mu, (\cdot)\big]~,
\ee
in terms of which we have
\begin{align}
D_1(D_1+i D_2)Z &= e_1^{\mu}D_{\mu}(D_1+iD_2)Z \cr
&= e_1^\mu\left({\cal D}_\mu(D_1+iD_2)Z - (\omega_\mu)^c_1  D_c Z - i (\omega_\mu)^c_2  D_c Z  \right)~.
\end{align}
The first term is now zero because of the BPS equations while the last term is non-zero due to the non-trivial spin-connection and we obtain
\begin{align}
D_1(D_1+i D_2)Z &=i D_3 Z ~.
\end{align}
One can do a similar calculation for the $D_2$ derivative and we obtain 
\be
(D_1-iD_2)(D_1+i D_2) Z = 2 i D_3Z\,.
\ee
Incorporating all these results and using the BPS equation to write the $D_0$ derivative in terms of the $D_3$ derivative, we therefore find that the equation of motion for the scalar field takes the form:
\be
-(D_3+i)^2 Z+D_3^2 Z + 2i D_3 Z- Z -[F_{12}, Z] + 2[Z,[Z,Z^{\dagger}] = 0~.
\ee
The first four terms add to zero while the last two terms can be rearranged to give
\be
\left [Z, F_{12} + 2  [Z,Z^{\dagger}] \right]= 0~,
\ee
which turns out to be identically true on account of the BPS equations. We have thus shown that the $\frac{1}{8}$-BPS equations imply the equation of motion for the scalar field, as expected -- thereby providing a consistency check on our BPS equations (\ref{finalBPS}).

Let us now turn to the equations of motion for the gauge field and the Bianchi identities:
\begin{align}
%\label{EoM}
D_a F^{ab}+ 2i \left( [Z^{\dagger}, D^bZ] - [D^b Z^{\dagger}, Z] \right) = 0~, \qquad
D_{[a}F_{bc]} = 0 ~.
\label{BIeqns}
\end{align}
There are eight equations here and, as we shall see, four of these will be satisfied identically due to the BPS equations. The remaining equations will impose additional differential equations that the gauge field and scalar field have to satisfy. 

As before, by splitting the $D_a$, in terms of the gauge and local Lorentz covariant derivative, the differential constraints that follow from the Bianchi identity are as follows:
\be
D_a F_{bc}= {\cal D}_a(\cdot) -(\omega_{a})^{d}_{\,\,b}F_{cd}-(\omega_{a})^{d}_{\,\,c}F_{db} = 0~. 
\ee
Using the explicit form of the spin connection and the algebraic constraints on the field strength that follow from the BPS equations, we find that the Bianchi identities are equivalent to the following equations:
\be
\begin{aligned}
D_0F_{12}-D_1 F_{02}+D_2F_{01} &= 0 ~,\quad
D_3F_{12}+D_1F_{02}-D_2F_{01} =0~, \\
\left( D_0 + D_3 \right) F_{01} + F_{02} &= 0 ~,\quad 
\left( D_0 + D_3 \right) F_{02} - F_{01} = 0~.
\label{allBIS}
\end{aligned}
\ee
Adding the two equations in the first row gives rise to the equation:
\be
(D_0 + D_3)F_{12} = 0\,,
\ee
which is automatically satisfied given the $\frac{1}{8}$-BPS equations (\ref{finalBPS}). The other combination will be dealt with later, along with an equation of motion. The two Bianchi identities in the second row of \eqref{allBIS} can be combined and written in the suggestive form:
\be
\label{dFone}
(D_0+D_3+i)(F_{01} - i F_{02}) = 0~.
\ee
Let us now turn to the equations of motion. As before, we first use the spin connection and the algebraic BPS equations involving the $F_{ab}$ to rewrite the equations of motion in the following form:
		\begin{align}
			\label{EOM1eightBPSF2}
			\left(  D_0 + D_3 \right) F_{01} + F_{02} &= 2i [Z^{\dagger}, (D_1 +i D_2) Z] - 2i [(D_1 - i D_2) Z^{\dagger}, Z]~, \\
			\label{EOM1eightBPSF3}
			\left( D_0 + D_3 \right) F_{02} - F_{01} &= 2 [Z^{\dagger}, (D_1 +i D_2) Z] + 2[(D_1 - i D_2) Z^{\dagger}, Z] ~,\\
					\label{EOM1eightBPSF1}
			D_1 F_{10} + D_2 F_{20} &= 2 i [ D_0 Z^{\dagger}, Z] - 2 i [Z^{\dagger} , D_0 Z] ~,\\
			\label{EOM1eightBPSF4}
			D_1 F_{10} + D_2 F_{20} + 2 F_{12} &=2 i [Z^{\dagger} , D_3 Z] - 2 i [ D_3 Z^{\dagger}, Z] ~.
		\end{align}
Using the BPS equations satisfied by the scalar field, the first two equations are completely equivalent to the complex differential condition in \eqref{dFone}, so, these do not lead to any new conditions. Taking the difference of the last two equations, we find 
\be
F_{12} = - i [\left( D_0 + D_3 \right) Z^{\dagger} , Z ] + i [Z^{\dagger} , \left( D_0 + D_3 \right)  Z ] =  2 [Z^{\dagger}, Z] ~,
\ee
which is identically satisfied on account of the BPS equations. 

After a little bit of algebra, the remaining equation of motion along with the leftover equation from the Bianchi identities can be combined into a single complex equation as follows: 
		\begin{align}
		\label{dFtwo}
			\left( D_1 + i  D_2 \right) \left( F_{01} - i F_{02} \right) = -4i [D_0 Z^{\dagger} , Z]
		\end{align}
Thus, the equations of motion and Bianchi identities impose four additional differential constraints on the field strengths and these are compactly written in terms of the two complex equations in \eqref{dFone} and \eqref{dFtwo}.

For the most part we shall focus on abelian solutions in the scalar sector in which we set the gauge fields to zero. For these solutions, the differential constraints on $F_{ab}$ which we derived in this section will not play any role. However, there are also defect-like solutions to the $\frac{1}{8}$-BPS equations involving only the gauge field in which we set the scalar field to zero. Such pure glue defects are outside the main focus of our work and we discuss the classical solutions and their charges briefly in Appendix \ref{SO6class}.

\section{Wobbling Strings}
\label{giantdefects}

So far we have seen that the static defects share supersymmetries with regular time dependent BPS solutions. In the rest of this work we will focus on the scalar sector, {\it i.e.} we set all the field strengths $F_{ab}=0$ and focus on non-trivial abelian scalar profiles. In this $\frac{1}{8}$-BPS scalar sector, the BPS equations take the simplified form:
\be
(D_0+D_3+i) \, Z  = 0~,\quad (D_1+i D_2) \, Z= 0~,
\label{Z1eqns}
\ee
where the covariant derivative $D_a$ is simply the vector field $E_a$ defined in \eqref{hopfvectorfields}. In terms of the coordinates on the sphere, these differential constraints are given by
\be
\begin{aligned}
\left( \pa_{\tau} + 2\pa_{\psi} + i \right) Z =0 \quad\text{and}\quad  \left( i\pa_{\vartheta} + (\csc\vartheta\pa_{\phi} - \cot\vartheta\pa_{\psi}) \right)Z = 0 ~.
\end{aligned}
\ee
Given the explicit differential operators, it is possible to write down a local Laurent series type solution that satisfies these differential constraints and it is given by
\be
\label{bdyprofile}
Z = \sum_{m,n}a_{m.n} e^{-i (m+n+1)\tau} \left( \cos\frac{\vartheta}{2}e^{ \frac{i}{2}(\psi+\phi)}\right)^m\,  \left( \sin\frac{\vartheta}{2}e^{\frac{i}{2}(\psi- \phi)}\right)^n~. 
\ee
While we will work with such explicit solutions in the following sections, we would like to characterize the most general solution to these equations in more general terms. To do so in a conceptually simple manner let us define
\be
\nu_0 = e^{i\tau}~, \quad \nu_1 = \cos\frac{\vartheta}{2}\, e^{\frac{i}{2}(\psi+\phi)}~, \quad \nu_2 = \sin\frac{\vartheta}{2}\, e^{\frac{i}{2}(\psi-\phi)}~.
\ee
Then we can write the general equation in a compact form as follows: 
\be
\label{ZS3}
Z\, \nu_0 = g\left(\frac{\nu_1}{\nu_0},\frac{\nu_2}{\nu_0} \right)~.
\ee
and it includes both regular as well as singular solutions depending on the analytic properties of the function $g(z_1, z_2)$. 

The $\nu_i$ are coordinates on a null cone in ${\mathbb C}^{1,2}$ that satisfies $-|\nu_0|^2 + |\nu_1|^2+|\nu_2|^2 = 0$. The $\nu_i$ can therefore be rescaled by a complex non-zero number $\lambda$ without affecting the fact that they are on a null-cone. In addition, we see that if this is accompanied by a rescaling of the complex field $Z$ by $\lambda^{-1}$, then, the equation defining the general solution to the BPS equations remains invariant. 

Our goal now is to characterize those solutions that correspond to a wobbling string in the gauge theory. By this we mean that the scalar field has a singularity at the location of the worldvolume of such a string. For any  given time $\tau$, this means that the scalar field should have a singularity along a one-dimensional path in $S^3$. To clarify this picture let us define the following scale-invariant variables:
\be
\label{zetadefn}
\zeta_0 = Z\, \nu_0~, \quad \zeta_1 = \frac{\nu_1}{\nu_0}~,\quad\text{and}\quad \zeta_2= \frac{\nu_2}{\nu_0}~. 
\ee
Thus time translation corresponds simply to scaling the $\nu_i$ for $i=1,2$ and $Z$ by a phase. 
Then, consider a solution of the BPS equations of the form
\be
\label{linkdefect}
\zeta_0\, F(\zeta_1, \zeta_2) - G(\zeta_1, \zeta_2) = 0~.
\ee
Here we assume both $F$ and $G$ to be analytic functions of its arguments. 
At the zeros of the function $F$, it is clear that the scalar field has a singularity. The locus of such points is a set ${\cal K}$ given by the intersection of 
\be
\label{Fwithsphere}
F(\zeta_1, \zeta_2)=0 \quad\text{and}\quad |\zeta_1|^2 + |\zeta_2|^2 = 1~.
\ee
Thus we find a simple characterization of the solutions to the $\frac{1}{8}$-BPS equations in the scalar sector that allows for a co-dimension two singularity in the solution for the $Z$-profile. The solutions to the equations in \eqref{Fwithsphere} are known to be algebraic links \cite{Hayden}. Thus at a given instant in time, the spatial configuration of the wobbling BPS string corresponds to a link in $S^3$. A particularly important class of solutions is given by choosing the function $F(z_1,z_2)$ to have a singularity structure at the origin. Then, it turns out that the topological type of the link stabilizes near the origin\footnote{We would like to thank T. Dimofte for clarifying this point.} and the intersection is known to give rise to a knot in $S^3$. For instance, for the choice of the function
\be
F(z_1, z_2) = z_1^p + z_2^q ~,
\ee
the solution to \eqref{Fwithsphere} is well known (see for instance  \cite{Milnor}) to be the torus knot $T_{p,q}$ (for $p,q \ge 2$ with $p$ and $q$ being coprime).  We note in passing that such knots have been studied in the context of topological string theory in \cite{Ooguri:1999bv, Diaconescu:2011xr}. Furthermore surface defects  have been proposed as a possible route to realize what mathematicians refer to as knot homologies \cite{Gukov:2007ck}. It would be interesting to see if our Hamiltonian approach might prove useful in this programme but we will not have more to say about these topics at this juncture. 

So far we have implicitly assumed that the solution for $\zeta_0$ be single-valued but it turns out that even this condition can be relaxed by considering a general solution of the BPS equations in terms of zeros of functions in the scale-invariant variables $H(\zeta_0, \zeta_1, \zeta_2)=0$. Since the variable $\zeta_0 = Z\, \nu_0$, and $Z$ is in general an eigenvalue of an $N\times N$ matrix, it follows that the holomorphic function can at most be of degree $N$ in $\zeta_0$. One can then factorize this polynomial in $\zeta_0$ and near each of its zeros, the general polynomial would factor into terms of the form in \eqref{linkdefect}. Remarkably we shall recover this general description of a wobbling string in a very natural way from the holographic description in terms of probe D3 branes in Section \ref{holostrings}. 

\subsection{Relation to Surface Operators}

In this section we relate the BPS strings we have studied to the codimension two defects in Euclidean space, focusing only on the scalar profiles. 
Let us start with conformally coupled scalar fields $Z$ and $\bar Z$ in ${\mathbb C}^2$ with complex coordinates $(z_1, z_2)$ and metric
\be
ds^2_{{\mathbb C}^2} = |dz_1|^2+|dz_2|^2~.
\ee
We now make the coordinate transformation $(z_1, z_2) = e^{\tau_E} (\cos\theta \, e^{i \phi_1}, \sin\theta \, e^{i\phi_2})$ in terms of which the metric takes the following form:
\be
ds^2_{{\mathbb C}^2} = e^{2\tau_E} (d\tau_E^2 + d\Omega_3^2)~. 
\ee
After Wick rotation $\tau_E = -i \tau$ this is therefore Weyl equivalent to the spacetime $S^3 \times {\mathbb R}$ with metric $-d\tau^2 + d\Omega_3^2$. The scalar fields $Z(z_i, \bar z_i)$ in ${\mathbb C}^2$ can be transformed into fields $Z(\tau, \theta, \phi_i)$ on $S^3 \times {\mathbb R}$ by using the fact that these scalars have Weyl weight $1$:
\be
Z'(x') = \Omega^{-1} Z(x)~,
\ee
where $\Omega$ is the Weyl factor that relates the two metrics $g'_{\mu\nu} = \Omega^2 g_{\mu\nu}$.

Let us start with the Gukov-Witten defect, which has the topology of a complex plane $\mathbb{C}\subset \mathbb{C}^2$. It is extended along the complex plane parametrized by $z_2$ and the scalar field $Z$ has a singular profile in the plane transverse to the defect, given by \cite{Gukov:2006jk}:
\be
Z_{{\mathbb C}^2}  = \frac{c}{z_1}~,
\ee
where we have indicated that this is the profile of the scalar field in the theory on $\mathbb{C}^2$. This corresponds to a conformal surface operator in ${\mathbb C}^2$. Let us transform it into a solution $Z_{S^3 \times {\mathbb R}} (\tau, \theta, \phi_i)$ on $S^3 \times {\mathbb R}$ spacetime by following the steps outlined above. Here the relevant Weyl factor is $\Omega = e^{-\tau_E}$. We have
\begin{align}
Z_{S^3 \times {\mathbb R}} (\tau, \theta, \phi_i) &= 
(e^{-\tau_E})^{-1} Z_{{\mathbb C}^2} (z_i, \bar z_i)\\ 
&= e^{\tau_E} \frac{c}{z_1} = \frac{c}{\cos\theta \, e^{i\phi_1}}.
\end{align}
As the final answer has no time dependence the configuration remains the same after Wick rotation. So we see that our half-BPS string in SYM on $S^3 \times {\mathbb R}$ which we studied in \eqref{class1def} maps to the conformal surface operator solution\footnote{The state-operator correspondence similarly maps the singular gauge profile $\underline{A}=\underline{\alpha} d\phi_1$ to the expected Gukov-Witten profile $\underline{A}_{\mathbb{C}^2} = -\frac{i}{2}\, \underline{\alpha}\, \left(\frac{dz_1}{z_1} - \frac{d\bar z_1}{\bar z_1}\right)$ in $\mathbb{R}^4$.} in SYM on ${\mathbb C}^2$. By considering the Gukov-Witten defect extended along $z_1$, we can similarly recover the BPS string solution with the singular profile in \eqref{Z1sintheta}. 

By following the same logic we can now relate our wobbling string solutions in the theory on $S^3 \times {\mathbb R}$ to configurations in ${\mathbb C}^2$. Recall from \eqref{ZS3} that our solutions are described by meromorphic functions of the form
\be
\label{S3timesRsolns}
Z_{S^3 \times {\mathbb R}}  = \frac{1}{\nu_0} g\left( \frac{\nu_1}{\nu_0}, \frac{\nu_2}{\nu_0}\right)
\ee
where $\nu_0 = e^{i\tau}= e^{-\tau_E}$, and $(\nu_1, \nu_2) = (\cos\theta \, e^{i \phi_1}, \sin\theta \, e^{i\phi_2})$. We already have that under the Weyl transformation,  $Z_{S^3 \times {\mathbb R}} = e^{\tau_E} \, Z_{{\mathbb C}^2} = \frac{1}{\nu_0} \, Z_{{\mathbb C}^2}$. Substituting these into (\ref{S3timesRsolns}) we arrive at:
\be
 Z_{{\mathbb C}^2} = g (z_1, z_2)~.
\ee
Thus we arrive at the conclusion that our $\frac{1}{8}$-BPS configurations in SYM on $S^3 \times {\mathbb R}$ translate into $Z = g(z_1, z_2)$ in the Euclidean theory on ${\mathbb R}^4$. Such surface defects preserving less than half of the supersymmetries have been described previously in \cite{Koh:2008kt}. In fact, the authors of \cite{Koh:2008kt}  also consider non-single valued configurations of the form $g(z_1, z_2) = (z_1z_2)^{-\frac{1}{2}}$ that involve fractional powers of the coordinates, accompanied by non-trivial gauge holonomy. We will discuss the energy and charges of such configurations in the following sections.  

\subsection{A New Variational Problem}

We are interested in studying properties of the BPS solutions (\ref{S3timesRsolns}) that may have singularities. Now, as we shall see in detail in this section, there are two potentially problematic issues in treating the singular solutions on par with the regular ones: (i) they do not belong to the same variational problem $\delta S =0$ and (ii) they have divergent energies, angular momenta and R-charges. 

In the following we will demonstrate that both these hurdles can be overcome by cutting off the spacetime arbitrarily close to the singularities of these solutions and adding appropriate boundary terms. In particular we will show that for a generic class of singular BPS solutions:
\begin{itemize}
\item It is possible to add boundary terms that make $\delta S=0$ as we vary along the space of solutions that include regular ones. This leaves a lot of ambiguity in the possible boundary terms.

\item Demanding that the global charges are rendered finite provides infinitely many conditions on the allowed set of boundary terms with $\delta S=0$ that essentially fixes them uniquely.
\end{itemize}

\subsubsection{On-shell Action and Boundary Terms}

Since we are in an abelian sector of the theory with a single complex scalar field $Z$, the theory reduces essentially to a conformally coupled complex scalar field on $ {\mathbb R}  \times S^3$, described by the Lagrangian:
\be
\label{cft-lagrangian}
{\cal L} = - \frac{1}{g_{YM}^2} \sqrt{-g} \left[ g^{\mu\nu} \partial_\mu Z \partial_\nu \bar Z + \bar Z \, Z \right]~.
\ee
We choose the line element on $S^3 \times {\mathbb R}$ to be
\be
\label{Rs3metric}
ds^2 = - d\tau^2 + (d\theta^2 + \cos^2\theta \, d\phi_1^2 + \sin^2\theta \, d\phi_2^2)~. 
\ee
The Lagrangian evaluated on the solutions:
\bea
Z = e^{i\tau} g\left(\widehat{\nu}_1, \widehat{\nu}_2 \right), ~~ \bar Z = e^{-i \tau} \bar g\left(\widehat{\bar \nu}_1, \widehat{\bar \nu}_2 \right)~,
\eea
with $\widehat \nu_i = \nu_i/\nu_0$, can be seen to be\footnote{In fact, this is true not only for the BPS solutions under discussion but also for a general solution of the equations of motion. It follows from the Virial theorem and the fact that the potential is quadratic.}
\bea
{\cal L}\Big{|}_{{\rm onshell}} = \frac{1}{2} \partial_\mu \left( Z \, \Pi_Z^\mu + \bar Z \Pi_{\bar Z}^\mu \right)
\eea
where we have introduced the conjugate momenta $\Pi_Z^\mu = \frac{\delta {\cal L}}{\delta \, (\partial_\mu Z)}$ and $\Pi_{\bar Z}^\mu = \frac{\delta {\cal L}}{\delta (\partial_\mu \bar Z)}$. 
These in turn can be written in terms of the function $g$ appearing in the solution as follows:
\be
\begin{aligned}
\label{mom-combos}
Z \, \Pi_Z^\theta + \bar Z \Pi_{\bar Z}^\theta &= -\frac{1}{g^2_{YM}} \cos\theta \sin\theta \left[ g \partial_\theta {\bar g} + \bar g \, \partial_\theta g \right]~, \\
Z \, \Pi_Z^{\phi_1} + \bar Z \Pi_{\bar Z}^{\phi_1} &= \frac{i}{g_{YM}^2} \tan\theta \, \left[ g \widehat {\bar{\nu}}_1 \partial_{\widehat {\bar \nu}_1}\bar g - \bar g \, \widehat \nu_1 \partial_{\widehat \nu_1} g \right]~, \\
Z \, \Pi_Z^{\phi_2} + \bar Z \Pi_{\bar Z}^{\phi_2} &= \frac{i}{g_{YM}^2} \cot\theta \, \left[ g \widehat {\bar{\nu}}_2 \partial_{\widehat {\bar \nu}_2}\bar g - \bar g \, \widehat \nu_2 \partial_{\widehat \nu_2} g \right] ~,\\
Z \, \Pi_Z^{\tau} + \bar Z \Pi_{\bar Z}^{\tau} &= \cos^2\theta \, (Z \, \Pi_Z^{\phi_1} + \bar Z \Pi_{\bar Z}^{\phi_1}) + \sin^2\theta \, (Z \, \Pi_Z^{\phi_2} + \bar Z \Pi_{\bar Z}^{\phi_2})~.
\end{aligned}
\ee
When the solutions are singular we propose to cut-off a region around (and arbitrarily close to) the singularities and add to the Lagrangian the boundary term 
\be
\label{Lbdy1}
{\cal L}_{bdy}^{(1)} = -\frac{1}{2} \widehat n_\mu \left( Z \, \Pi_Z^\mu + \bar Z \Pi_{\bar Z}^\mu \right)~,
\ee 
where $\widehat n_\mu$ is the unit outward normal to the boundary. This ensures that the Lagrangian evaluates to zero for all solutions -- regular as well as the  singular ones -- thus making all the solutions belong to the same variational problem. 

One can check that the solutions satisfy the following two constraints: 
\be
\begin{aligned}
\label{cft-constraints}
\Pi^\theta_Z + i \, \cos\theta \sin\theta \, (\Pi_Z^{\phi_1} - \Pi_Z^{\phi_2}) &=0~, \cr
( \Pi_Z^{\phi_1} \, \cos^2\theta + \Pi_Z^{\phi_2} \, \sin^2\theta ) -  \Pi_Z^\tau - \frac{i}{2} \cos\theta \sin\theta ~ \bar Z &= 0~.
\end{aligned}
\ee
along with their complex conjugates. These relations are identically satisfied on account of the BPS equations.

There are some interesting subclasses of solutions for which some of the combinations in (\ref{mom-combos}) vanish. In particular
\begin{enumerate}
\item For $g(\widehat \nu_1, \widehat \nu_2)$ is homogenous in $(\widehat \nu_1, \widehat \nu_2)$: 
\bea
g(\lambda \, \widehat \nu_1, \lambda \widehat \nu_2) = \lambda^p g(\widehat \nu_1, \widehat \nu_2) \iff Z \,\Pi_Z^{\tau} + \bar Z \Pi_{\bar Z}^{\tau}=0~.
\eea
\item For $g(\widehat \nu_1, \widehat \nu_2)$ is homogeneous in $\widehat \nu_1$:
\bea
g(\lambda \widehat \nu_1, \widehat \nu_2) = \lambda^{p_1} g(\widehat \nu_1, \widehat \nu_2) \iff Z \, \Pi_Z^{\phi_1} + \bar Z \Pi_{\bar Z}^{\phi_1}=0~.
\eea
\item For $g(\widehat \nu_1, \widehat \nu_2)$ is homogeneous in $\widehat \nu_2$:
\bea
g(\widehat \nu_1, \lambda \widehat \nu_2) = \lambda^{p_2} g(\widehat \nu_1, \widehat \nu_2) \iff Z \, \Pi_Z^{\phi_2} + \bar Z \Pi_{\bar Z}^{\phi_2}=0~.
\eea
\end{enumerate}
We will restrict to the class for which $Z \,\Pi_Z^{\mu} + \bar Z \, \Pi_{\bar Z}^{\mu}$ vanishes for $\mu = \tau, \phi_1, \phi_2$. This requires that the function $g(\widehat \nu_1, \widehat \nu_2)$ is homogeneous in each of its arguments separately. This restricts $g$ to be a monomial: $g(\widehat \nu_1, \widehat \nu_2) = c_{mn} \widehat \nu_1^m \, \widehat \nu_2^n$. These are natural (time-dependent) generalizations of the simple (static) surface defect. 
In terms of the coordinates on $\mathbb{R}\times S^3$ in \eqref{Rs3metric}, the scalar profile takes the following form:
\be
\label{solution1}
Z = r_0 \, e^{i \, (\xi_0 - \tau)} \, \left(\cos\theta \, e^{i \, (\phi_1 - \tau) }\right)^m  \left(\sin\theta \, e^{i \, (\phi_2 - \tau)}\right)^n 
\ee
These solutions are specified by one complex parameter $\eta = r_0 \, e^{i \, \xi_0}$ and two real parameters $(m, n)$. If we demand that the complex field $Z$ is single valued and periodic under $\phi_i \rightarrow \phi_i + 2\pi$, this allows only integer $(m,n)$ pairs. One may also consider $(m,n)$ to be rationals if we relax these conditions and we will find that even these lead to finite energies and charges. We list a few special cases within this monomial class:
\begin{enumerate}
\item When $m+n+1=0$ the solution becomes static. The 1/2-BPS ``conformal" defects correspond to the special case of $(m,n) = (0,-1)$ or $(-1,0)$ and these preserve an $so(2,2) \subset so(2,4) $ subalgebra of the theory. 

\item The duals of 1/2-BPS ``round" dual-giants correspond to $(m,n) =(0,0)$ and these solutions respect an $so(4)$ subalgebra. 

\item If we set one of either $m$ or $n$ to zero, the BPS string maps to the surface defects with wild ramification in $\mathbb{R}^4$ that were defined in \cite{Witten:2007td}. 
\end{enumerate}

\noindent
The solution (\ref{solution1}) is singular at $\theta =0$ ($\theta = \pi/2$) for negative values of $n$ ($m$). 
We now work with the monomial solution and exhibit in this example the general features of the variational problem. The Lagrangian density (\ref{cft-lagrangian}) evaluated on (\ref{solution1}) gives:
\be
\label{Ldensity}
{\cal L} (r_0, m,n) =\frac{2 r_0^2}{g_{YM}^2} \cos^{2m}\theta \, \sin^{2n} \theta\left[(m+n)(m+n+1) \, \cos\theta \, \sin\theta  -m^2 \tan\theta - n^2 \cot\theta  \right]  
\ee
which can be written as 
\be
\label{intL}
{\cal L} (r_0, m,n) = \frac{d}{d\theta}\left(\frac{r_0^2}{g_{YM}^2} \cos^{2m+1}\theta \, \sin^{2n+1}\theta \, \left( m \, \tan\theta - n \, \cot\theta \right)\right)~.
\ee
Therefore the Lagrangian density (\ref{Ldensity}) when integrated over $\theta$ between $0$ and $\pi/2$ vanishes for non-negative $m$ and $n$, which in turn means that we have $\delta S =0$ when we vary along the space of solutions (\ref{solution1}) by changing the parameters $0 \le r_0, m, n < \infty$. But for $m < 0$ or $n< 0$ its integral diverges. In particular for $n <0$ ($m<0$) the singularities come from $\theta =0$ ($\theta = \frac{\pi}{2}$) region. 
As we have discussed in generality, to include these BPS defects into the same variational problem we propose to cut-off the region around the surface defect and add the boundary term in \eqref{Lbdy1}. For the monomial solutions with $n<0$, this corresponds to adding 
\be
\label{cft-bdy1}
{\cal L}^{(1)}_{bdy, 0+} =  \frac{1}{2} (Z \, \Pi_Z^\theta + \bar Z \, \Pi_{\bar Z}^\theta)
\ee
at $\theta = 0+\epsilon$, and for those solutions with $m<0$
\be
\label{cft-bdy2}
{\cal L}^{(1)}_{bdy, \frac{\pi}{2}-} = - \frac{1}{2} (Z \, \Pi_Z^\theta + \bar Z \, \Pi_{\bar Z}^\theta)
\ee
at $\theta = \frac{\pi}{2} - \epsilon$. However demanding $\delta S =0$ still leaves a lot of freedom in the possible boundary terms. For example one may add any term that is proportional to the constraints (\ref{cft-constraints}) such as:
\be
\label{Lbdy2}
f_{\cal C}(Z, \Pi^\mu_Z) \, {\cal C}(Z, \Pi^\mu_Z) + c.c
\ee
where ${\cal C}(Z, \Pi^\mu_Z)$ is one of the constraints in (\ref{cft-constraints}) as these terms would vanish identically on-shell. We will exploit this freedom to regularize the energy and other charges.

\subsubsection*{Comments on the variational problem}

 The boundary terms we have added are novel and we pause briefly to outline the approach we have followed in finding them. The change in the action under a general variation is as follows:
\begin{align}
\delta S 
&= \int d^4x \, \left({\bf EoM}_Z \, \delta Z + {\bf EoM}_{\bar Z} \, \delta {\bar Z} \right) +  \frac{1}{2} \int d^4x \, \partial_\mu \left(\Pi_Z^\mu \, \delta Z + \Pi_{\bar Z}^\mu \, \delta {\bar Z}  -  Z \, \delta\Pi_Z^\mu - \bar Z \, \delta \Pi_{\bar Z}^\mu \right) \cr 
&=\int d^4x \, \left({\bf EoM}_Z \, \delta Z + {\bf EoM}_{\bar Z} \, \delta {\bar Z} \right) + \frac{1}{2} \int d\Sigma^{(3)}\hat n_\mu \left(\Pi_Z^\mu \, \delta Z + \Pi_{\bar Z}^\mu \, \delta {\bar Z}  -  Z \, \delta\Pi_Z^\mu - \bar Z \, \delta \Pi_{\bar Z}^\mu \right)  ~,
\end{align}
where ${\bf EoM}_Z$ is proportional to the equation of motion of $Z$. Since the classical solutions have to belong to the configuration space of the theory and should have $\delta S =0$ --  the configuration space is constrained to be such that the boundary term in $\delta S$ vanishes:
\begin{equation}
\int d\Sigma^{(3)}\hat n_\mu \left(\Pi_Z^\mu \, \delta Z + \Pi_{\bar Z}^\mu \, \delta {\bar Z}  -  Z \, \delta\Pi_Z^\mu - \bar Z \, \delta \Pi_{\bar Z}^\mu \right)
 =0~.
\end{equation}
This condition translates into the specification of boundary conditions that any configuration has to satisfy. Because we are working on $S^3 \times {\mathbb R}$ for any time-like boundary the boundary would have the topology of $\Sigma_2 \times {\mathbb R}$ for some compact space $\Sigma_2$. For instance if the boundary is at $\theta = \theta_0$ then $\sigma_2$ is a $T^2$ with metric $\cos^2\theta_0 \, d\phi_1^2 + \sin^2 \theta_0 \, d\phi_2^2$.  We will work with this boundary from now on. 

In this case (the boundary given by constant $\theta$) the vector $\hat n_\mu = (0,1,0,0)$ and the vanishing of the boundary term 
\begin{equation}
\int_{-\infty}^\infty d\tau \int_0^{2\pi} \int_0^{2\pi}d\phi_1\, d\phi_2 ~ \left( \Pi_Z^\theta \, \delta Z + \Pi_{\bar Z}^\theta \, \delta {\bar Z}  -  Z \, \delta\Pi_Z^\theta - \bar Z \, \delta \Pi_{\bar Z}^\theta \right)~,
\end{equation}
is ensured by the following condition at the boundary $\theta = \theta_0$: 
\begin{equation}
\label{bcs-one}
\Pi_Z^\theta \, \delta Z + \Pi_{\bar Z}^\theta \, \delta {\bar Z}  -  Z \, \delta\Pi_Z^\theta - \bar Z \, \delta \Pi_{\bar Z}^\theta=\partial_{\phi_1} f^{\phi_1} + \partial_{\phi_2} f^{\phi_2}~.
\end{equation}
Here $f_{\phi_1}$ and $f_{\phi_2}$ are some (real) functions that are periodic under $\phi_i \rightarrow \phi_i + 2\pi$. We will take this to be  the boundary condition on our general configurations. Note that these are neither Dirichlet ($\delta Z =\delta {\bar Z} =0$) type nor Neumann type ($\delta\Pi_Z^\theta =\delta\Pi_{\bar Z}^\theta=0$) -- but some mixed ones. 

We are interested in ensuring that our BPS solutions are part of the solution space. This requires that our BPS configuration satisfy the boundary conditions (\ref{bcs-one}). To check this we need to verify that when we take the configurations to be BPS ones and the variations to be along the space of solutions the boundary conditions have to be satisfied. If true this will ensure that the configuration space is at least as big as the space of all configurations that approach (sufficiently fast) one of the BPS solutions near the boundary. 

It is now easy to verify using the BPS equations in \eqref{cft-constraints} and its complex conjugate, that when we are moving along the BPS solutions space the choice of functions is given by:
\begin{equation}
\label{bpsfs}
f^{\phi_1} = \frac{i}{g^2_{YM}} \sin^2\theta\, ({\bar Z} \, \delta Z - Z \, \delta {\bar Z}), ~~~ f^{\phi_2} = -\frac{i}{g^2_{YM}} \cos^2\theta \, ({\bar Z} \, \delta Z - Z \, \delta {\bar Z})~.
\end{equation}
For this choice of ($f^{\phi_1}, f^{\phi_2}$) our boundary conditions (\ref{bcs-one}) become:
{\small
\begin{equation}
\label{bps-bcs2}
Z^2 \, \delta \! \left[ \frac{1}{Z} \left( \Pi^\theta_Z + i \, \cos\theta \sin\theta \, (\Pi_Z^{\phi_1} - \Pi_Z^{\phi_2}) \right) \right] 
+ {\bar Z}^2 \, \delta \! \left[ \frac{1}{\bar Z} \left( \Pi^\theta_{\bar Z} - i \, \cos\theta \sin\theta \, (\Pi_{\bar Z}^{\phi_1} - \Pi_{\bar Z}^{\phi_2}) \right) \right] =0 ~.
\end{equation}
}
Symbolically, this can be written as 
$
g_1 ~\delta {\cal C} + \delta g_2~ {\cal C} = 0~,
$
where $g_i$ are arbitrary functions on phase space. This essentially shows that a consistent variational problem is obtained if the boundary conditions are such that the BPS constraint ${\cal C} =0$ is obeyed and the variations  are such that they too satisfy $\delta C=0$ near the boundary.

We will therefore define our configuration space to consist of all configurations that approach (sufficiently fast) one of the BPS solutions near the boundary. This means that the boundary condition we impose is (\ref{bps-bcs2}) at the boundary. The BPS string configurations that we work with in this work are singular at either $\theta =0$ or $\theta = \pi/2$ and we excised out the singular region to create a boundary and the divergence of the action comes from the singular region. We have also shown that for singular BPS configurations the action is rendered finite by our boundary term. It should be clear that for any configuration which is regular and smooth (both fields and their gradients are finite) and approach one of the BPS configurations at the boundary its action (actually the Lagrangian density integrated over the space coordinates but not time) will be finite. So far we have only worked with the first boundary term that we added in \eqref{Lbdy1}. It is also easy to state what the boundary conditions should be when we have the additional boundary terms in \eqref{Lbdy2}: we just have to keep adding their variations to the l.h.s of (\ref{bcs-one}). Since the new terms in the boundary Lagrangian are constraints that vanish for BPS configurations we do not have to change the definition of the configuration space.

\subsection{Regularization of Charges}
\label{YMregular}
The theory in (\ref{cft-lagrangian}) is invariant under translations in $\tau, \phi_1, \phi_2$ and the global $u(1)$ R-symmetry, and we refer to the corresponding conserved charges as $E, S_1, S_2$ and $J$ respectively. The Hamiltonian density is given by
\be
{\cal E} = Z \frac{\delta {\cal L}}{\delta \dot Z} + \bar Z \frac{\delta {\cal L}}{\delta \dot {\bar Z}} - {\cal L}~,
\ee
and by evaluating this on the solutions we obtain
\be
\label{cft-energyD}
{\cal E} (r_0, m,n) =\frac{ 2r_0^2}{g_{YM}^2} \, \cos^{2m}\theta \, \sin^{2n} \theta \, \left[ (m+n+1) \, \cos\theta \, \sin\theta + m^2 \, \tan\theta + n^2 \cot \theta \right]
\ee
For $m,n\ge 0$ this gives the energy $E = 4 \pi^2 \,\int_0^{\frac{\pi}{2}} d\theta ~ {\cal E}$ :
%(
\be
\begin{aligned}
\label{wso-energy}
E(r_0, m,n) &= 4\pi^2 \, \frac{2r_0^2}{g_{YM}^2} \, \frac{(m+n+1)}{2} \frac{\Gamma(1+m)\Gamma(1+n)}{\Gamma(m+n+1)} \, \cr &=  \frac{4\pi^2}{g_{YM}^2} \, r_0^2 \, (m+n+1)^2 \, B(m+1,n+1)~,
\end{aligned}
\ee
where $B(a,b)$ is the Euler Beta function. The other charges for $m,n \ge 0$ are computed from the stress tensor (see Appendix \ref{Tmunu}) and we obtain
\be
\begin{aligned}
J(r_0,m,n) &= \frac{4\pi^2 \, r_0^2}{g_{YM}^2} \, (m+n+1) \, B(m+1,n+1) , \cr 
S_1(r_0,m,n) &= -\frac{4\pi^2 \, r_0^2}{g_{YM}^2} \, (m+n+1) m \, B(m+1,n+1) , \,\cr  
S_2(r_0,m,n) &= -\frac{4\pi^2 \, r_0^2}{g_{YM}^2} \, (m+n+1) \,n \, B(m+1,n+1) ~.
\end{aligned}
\label{ES1S2}
\ee
%wso-energy, ES1S2
It is easy to check that the charges satisfy the linear relation: $E + S_1 + S_2 =J$. 

For negative $m$ or $n$ the energy ($E$) and the other charges $(S_1, S_2, J)$ are all divergent. Also we  need to take the contributions of the boundary terms into account. We will show below that -- after subtracting the divergences -- this answer (\ref{wso-energy}) so far valid for non-negative $(m,n)$ can be treated as the answer for other values of $(m,n)$ when it is finite. For this let us find the nature of these singularities first. 

The energy density (\ref{cft-energyD}) can be rewritten in the form:
\be
\begin{aligned}
\label{cft-energyDv2}
& \frac{g^2_{YM}}{2 \, r_0^2} {\cal E} (r_0, m,n) = \cos^{2m}\theta \, \sin^{2n} \theta \, \left[ (m+n+1) \, \cos\theta \, \sin\theta + m^2 \, \tan\theta + n^2 \cot \theta \right] \cr
&= (m+n+1)^2 \cos^{2m+1}\theta \, \sin^{2n+1} \theta -\frac{d}{d\theta} \left[ \frac{1}{2}\cos^{2m}\theta \, \sin^{2n} \theta \, ( m \, \sin^2\theta - n \, \cos^2\theta) \right]~. \cr 
\end{aligned}
\ee
Moving the total derivative term to the l.h.s and integrating over $\theta$ gives
\begin{subequations}
\begin{align}
\label{bulkintegralv1line1}
{\cal I} &= \frac{g^2_{YM}}{2 \, r_0^2} \int d\theta ~ {\cal E} (r_0, m,n)
+ \frac{1}{2}\cos^{2m}\theta \, \sin^{2n} \theta \, ( m \, \sin^2\theta - n \, \cos^2\theta)\\
&= (m+n+1)^2 \, \int d\theta \, \cos^{2m+1}\theta \, \sin^{2n+1} \theta~.
\label{bulkintegralv1}
\end{align}
\end{subequations}

When the parameters $(m,n)$ take negative values, one can check that the energy is divergent. The second term on the r.h.s of \eqref{bulkintegralv1line1} is responsible for the leading divergence in the bulk contribution to the energy as, near $\theta\rightarrow 0$, it diverges as $\theta^{2n}$ for $n<0$. A similar power law divergence appears for $m<0$ as $\theta\rightarrow \frac{\pi}{2}$. In fact, these divergences get cancelled by the energy contributions from the boundary terms (\ref{cft-bdy1}) and \eqref{cft-bdy2} respectively. 

We will consider the cases $(m>0 , n<0)$ and $(m<0, n>0)$ differently in what follows. The integral in \eqref{bulkintegralv1} can be evaluated as
\begin{equation}
{\cal I}=
  \begin{cases}
    - \frac{(m+n+1)^2}{2 \, (m+1) } \cos^{2m+2} \theta \, F(1+m, -n, 2+m, \cos^2\theta)~, & \text{if $m\ge 0$ and $n<0$}.\\
  \phantom{-} \frac{(m+n+1)^2}{2 \, (n+1) } \sin^{2n+2} \theta \, F(1+n, -m, 2+n, \sin^2\theta)~, & \text{if $m<0$ and $n\ge 0$}.
  \end{cases}
  \label{mnbulkcases}
\end{equation}
Here $F$ denotes the hypergeometric function ${}_2F_{1}(a,b,c;z)$. 
Now, for $(m>0, n<0)$, it is clear that the defect is located at $\theta=0$. As a consequence, it is only possible to cancel (divergent) contributions to the energy density that arise from $\theta=0$, by adding suitable boundary term at $\theta=0$. On the other hand, it would be important to ensure that there is no contribution at all from the other end of the bulk integration, namely at $\theta=\frac{\pi}{2}$. The choice we have made in \eqref{mnbulkcases} ensures that this is so. 

The derivative of either of these terms leads to the same energy density and so these differ by a constant that only depends on $(m,n)$ and is independent of $\theta$. By explicitly evaluating a few cases it becomes clear that the difference is precisely given by $E(r_0,m,n)$ given in \eqref{wso-energy}, but analytically continued to negative values of either $n$ or $m$, leading to finite results. This will prove to be important when the regularized energies are calculated. 

With the choices that we have made, for the $(m>0, n<0)$ cases, the contribution of ${\cal I}$ to the energy leads to power law (and in some special cases logarithmic) divergences near $\theta=0$. In order to cancel these divergences we add boundary terms of the form 
\bea
L'_{bdy} = f(m,n,\theta)\frac{i}{g^2_{YM}} \, (\Phi \, {\cal C} - \bar \Phi \, \bar {\cal C})~,
\eea
where ${\cal C}$ is the constraint
\be
{\cal C} = ( \Pi^{\phi_1} \, \cos^2\theta + \Pi^{\phi_2} \, \sin^2\theta ) -  \Pi^\tau + \frac{i}{2} \cos\theta \sin\theta ~ \Phi~,
\ee 
and $\bar {\cal C}$ its complex conjugate and $f(m,n,\theta)$ is a real function of $Z, \bar Z, \Pi^{\phi_i}_Z, \Pi^{\phi_i}_{\bar Z}$ obtained by the following replacements:
\be
\label{mnsubs}
m \longrightarrow -i\, 2g_{YM}^2\, \cot\theta \left( \frac{\Pi^{\phi_1}_Z}{\bar Z} -  \frac{\Pi^{\phi_1}_{\bar Z}}{Z} \right), ~~ n \longrightarrow -i \, 2g_{YM}^2\, \tan\theta \left( \frac{\Pi^{\phi_2}_Z}{\bar Z} -  \frac{\Pi^{\phi_2}_{\bar Z}}{Z} \right)~.
\ee
It is clear that such a term will not alter the property that the on-shell action vanishes since it is proportional to the phase space constraints. However, it does contribute to the energy a term proportional to $\frac{2 \, r_0^2}{g^2_{YM}}$ times
\bea
\frac{1}{2} (m+n+1) \, \cos^{2m+1} \theta \, \sin^{2n+1}\theta \, f(m,n,\theta)
\eea
to the energy. At first glance this does not appear to be of the same form as the bulk integral in \eqref{mnbulkcases}. However by using the Pfaff transformation 
\be
F(a,b,c;z) = (1-z)^{-b} \, F\big(c-a,b,c,; \frac{z}{z-1}\big)~, 
\ee
we can write the terms on the r.h.s in \eqref{mnbulkcases} as:
\be
{\cal I } = 
  \begin{cases}
- \frac{(m+n+1)^2}{2 \, (m+1) } \cos^{2m+2} \theta \, \sin^{2n}\theta \, F(1, -n, 2+m, -\cot^2\theta)~, & \text{if $m\ge 0$ and $n<0$}.\\
\phantom{-}\frac{(m+n+1)^2}{2 \, (n+1) } \cos^{2m} \theta \,\sin^{2n+2} \theta \, F(1, -m, 2+n, -\tan^2\theta)~, & \text{if $m< 0$ and $n\ge 0$}.
\end{cases}
\ee
At this point, there are two possibilities for the coefficient $f(m,n,\theta)$. One possibility is to add the boundary term that exactly cancels the bulk divergence; this would imply that for all $(m,n)$ the energy of these defect contributions is {\it exactly} zero and the same is true for all the other conserved charges. From various points of view, we feel that this is not a reasonable outcome and we make the alternative choice in which the boundary Lagrangian is 
\be
f(m,n,\theta) = 
\begin{cases}
-\frac{m+n+1}{ n+1 } \tan\theta \, F(1, -m, 2+n, -\tan^2\theta)~, & \text{if $m\ge 0$ and $n<0$}.\\
\phantom{-}\frac{m+n+1}{ m+1 } \cot\theta \, F(1, -n, 2+m, -\cot^2\theta)~, & \text{if $m<0$ and $n\ge 0$}.
\end{cases}
\label{fmboundary}
\ee
One reason for making this choice is that the hypergeometric terms have a sensible power series expansion near the respective boundaries. Combining now the bulk and boundary contributions we see that the energy of the wobbling string takes is simply the analytic continuation of the energy $E(r_0, m,n)$ to negative values of either $n$ or $m$. 

There is another closely related reason for choosing the first (second) of (\ref{fmboundary}) in the boundary term at $\theta =0$ ($\theta = \frac{\pi}{2}$). Suppose we move through the space of polynomial solutions from a wobbling string at $\theta =0$ to a regular one at $\theta =0$. This can be achieved by simply tuning the sign of $n$ from negative to positive. Then the energy ceases to be singular. But if we use the second boundary term at $\theta =0$ this would cancel the energy exactly irrespective the sign of $n$ and would make the energy of the smooth solutions zero. But we know that the energies of smooth solutions are finite -- as for instance they can be compared directly with the corresponding finite energies of wobbling dual-giants.

We conclude this section by examining the status of the other charges $(S_1, S_2, J)$ and the linear relation $E = S_1 + S_2 +J$. It turns out that the same set of boundary terms that allowed us to include the defects to the class of dual-giant like solutions also regularize these charges, and maintain the linear BPS relation between the charges valid for these solutions as well.

Even though we appear to have added infinitely many counter terms (as seen by expanding the hypergeometric function as a power series) to render the energy finite, it must be pointed out that for a given $(m,n)$ only finitely many of the boundary terms are responsible for cancelling the divergences.

Our prescription to subtract away the (coordinate-dependent) divergences in the charges of the BPS string solutions may appear to be ad hoc. However, it will become clear from the dual holographic analysis that our prescription here is nothing more than the standard UV renormalization. In particular our prescription should be treated on the same footing as the renormalization of the Euclidean action for Wilson line or surface defect operators that gives rise to finite expectation values of those non-local operators (see for instance \cite{Drukker:2005kx, Gomis:2007fi, Drukker:2008wr}).

\subsection{Some Examples}

Our analysis in the previous section shows that it is possible to add a boundary term to the classical action such that all classical singular profiles described by \eqref{solution1} are solutions to the same variational problem. Further we showed that it is possible to add extra boundary terms proportional to the phase space constraints such that the energy turns out to be finite in a large number of cases. In this section we focus on a few simple cases in which the boundary Lagrangian simplifies considerably and that illustrate the general ideas that we have discussed so far.

\subsubsection{The static case $m+n+1=0$:}

In this case, it is straightforward to see from \eqref{solution1} that the solution is time independent. We focus on the case in which $m>0$ and $n<0$ and the profile for the scalar field takes the form
\be
Z = r_0 \, e^{i\xi_0} (\cos\theta \, e^{i\phi_1})^{-n-1}(\sin\theta \, e^{i\phi_2})^{n}~.
\ee
The on-shell momenta $\Pi_Z^{\tau}$ vanishes and the requirement that the energy be finite reduces to requiring that  the on-shell action be finite. In this case, the function $f(m,n,\theta)$ vanishes and we obtain the simple boundary term in \eqref{cft-bdy1}:
\be
L_{bdy} =  \frac{1}{2} (Z \, \Pi_Z^\theta + \bar Z \, \Pi_{\bar Z}^\theta)~,
\ee
where this is evaluated near $\theta=0$. A similar result holds for the case with $m<0$ and $n>0$, in which the boundary term is added at $\theta=\frac{\pi}{2}$. The simple $\frac{1}{2}$-BPS defects, which correspond to the $(0,-1)$ and $(-1,0)$ cases fall into this category and we find that the on-shell action and energy vanish with this boundary term\footnote{It is possible to put the ${\mathcal N}=4$ gauge theory on different backgrounds such as $AdS_4$, $AdS_3\times S^1$ \cite{Drukker:2008wr}, $dS_4$ \cite{Jensen:2018rxu}, etc. The corresponding half-BPS conformal defect can be obtained in each of these cases using the appropriate Weyl transformation. The Killing energies computed in the various cases are physically distinct.}.

\subsubsection{The strings of type $(0,n)$}
\label{0ndefects}

These are the time dependent solutions of the form
\be
Z = r_0 \, e^{i \, (\xi_0 - \tau)} \,  \left(\sin\theta \, e^{i \, (\phi_2 - \tau)}\right)^n ~,
\ee
with $n<0$. One can check that the function $f(0,n,\theta)$ simply reduces to unity and the boundary action takes a particularly simple form:
\be
\label{0nbdyaction}
L_{bdy} + L'_{bdy} = \frac{1}{2} (Z \, \Pi_Z^\theta + \bar Z \, \Pi_{\bar Z}^\theta) + \frac{i}{2} \tan\theta \, (Z \, {\cal C} - \bar Z \, \bar {\cal C})~,
\ee
where ${\cal C}$ is the constraint in \eqref{cft-constraints}. The renormalized energy in this case turns out to be
\be
\label{E0n}
\frac{E_{0,n}}{4\pi^2} = \frac{(n+1) }{g_{YM}^2}\, r_0^2~.
\ee
Note that the energy is negative for negative values of $n$.

\subsubsection{The strings of type $(-n,n)$}
\label{minusnndefects}

Here we again consider cases with $n<0$. In this case there is no simplification that occurs in the boundary term and one has both the boundary terms:
\be
L_{bdy} + L'_{bdy} = \frac{1}{2} (Z \, \Pi_Z^\theta + \bar Z \, \Pi_{\bar Z}^\theta) + \frac{i}{2}  \, (Z \, {\cal C} - \bar Z \, \bar {\cal C})\, f(-n,n,\theta)~.
\ee
This is an interesting case and for fractional (rational) values of $n$, these give finite results. For instance, for any positive integer $p$, we have
\be
\label{E-nn}
\left.\frac{ E_{-n,n}}{4\pi^2}\right|_{n=-\frac{(2p+1)}{2}} =(-1)^{p} \frac{\pi\, (2p+1)}{2g_{YM}^2}\, r_0^2~. 
\ee

\subsection{Summary}

We end this section with some general remarks about the status of these singular solutions and their renormalized finite energies. First of all we have given a characterization of what we have termed wobbling string solutions in terms of holomorphic functions that intersect the 3-sphere. For such $\frac{1}{8}$-BPS solutions we showed that it is possible to define a consistent variational problem that includes all those solutions in a consistent manner, and such that the on-shell action is zero. We will have more to say about the general features of these solutions in Section \ref{holostrings} where we discuss them from a holographic point of view. 

We focused on solutions of the monomial type functions given in \eqref{solution1}. For the case that $(m,n)$ are both positive, these are well studied in the literature and give rise to finite energies and global charges, as can be seen from the explicit expressions in \eqref{wso-energy} and \eqref{ES1S2}. We then went on to consider solutions for which one of $m$ or $n$ is negative (which correspond to particularly simple wobbling string solutions) and showed that it was possible to add further boundary terms such that the renormalized energy and charges were simply an analytic continuation of the formulae obtained for the cases with $(m,n)$ positive. For particular sub-cases we observed that the energies can be positive or negative and the interpretation of this is unclear at this juncture. It is likely that these could be interpreted as a Casimir energies of the worldvolume theories on the stringy defects. 

Let us now discuss the limitations of our analysis. We recall that when we studied the possible boundary terms that could be added there were two natural possibilities: one led to a cancellation of {\it all} divergences in the charges (the on-shell action vanishes for all BPS solutions) but at the same time also cancelled the finite parts such that the energy and global charges turned out to be zero for all monomial solutions. This was considered to be an unphysical choice\footnote{As alluded to previously, one could consider cutting off even the regular monomial solutions away from $\theta=0$ and adding the same set of same boundary terms as for the singular solutions. This particular choice would lead to regular solutions with zero energy and charges, which is unphysical.} and we instead opted for the choice in which the energy and global charges were analytic continuations of the corresponding results for the cases in which both $(m,n)$ are positive real numbers. For those cases in which the analytic continuation of the energy and charges lead to finite values, our proposal gives a complete renormalization prescription. In fact, in Section \ref{wobblers} we will interpret this as holographic renormalization and compute the subleading (in 't Hooft coupling) corrections to the energy and charges by considering the renormalization of the probe D3-brane theory. 

However, it is evident that there are choices of $(m,n)$ for which the analytic continuation of the charges does not lead to finite results. For instance, for the case of $m+n=0$, the analytic continuation of the energy in \eqref{wso-energy} to negative integer values of $n$ leads to divergent results. On  a careful examination, it turns out that, while the coordinate dependent power law divergences do cancel, there are additional singularities that could be interpreted as logarithmic singularities. Similar divergences also appear for $m$ and $n$ both sufficiently negative. As it stands, for those cases for which the analytic continuation does not lead to a finite result, we do not have a proposal for regularizing the BPS string solutions, and further analysis of such solutions is beyond the scope of this work. Our primary goal in going through the regularization procedure was to demonstrate finiteness of energy and charges for a large class of monomial solutions.

\section{D3-brane Probes in \texorpdfstring{$AdS_5\times S^5$}{ads5s5}}
\label{D3probes}

We now turn to a holographic description of the wobbling strings by considering probe D3-brane solutions in global $AdS_5\times S^5$.  
In the Euclidean context, in \cite{Drukker:2008wr, Koh:2008kt}, the holographic duals of surface operators that preserve some fraction of the supersymmetry with topology $\mathbb{R}^{2}\subset \mathbb{R}^{4}$ were shown to be described by probe D3-branes ending on the boundary in a two dimensional surface. In this section we perform the analogous calculation in global coordinates. For the half-BPS string, we will find that the probe brane ends on the boundary on a surface with topology $\mathbb{R}\times S^1 \subset \mathbb{R}\times S^3$. 

\subsection{The closed string background}

We begin with a brief review of the geometry of the bulk background and we begin by defining the following complex coordinates:\footnote{These complex coordinates of the ambient space will turn out to be useful when we discuss the more general probe branes as zeros of holomorphic functions.} 
\be
(\Phi_0, \Phi_1, \Phi_2, Z_1, Z_2, Z_3)  \in \mathbb{C}^{1,2}\times \mathbb{C}^3~.
\ee
The $AdS_5\times S^5$ background is defined as the following locus in this ambient space:
\be
-|\Phi_0|^2 + |\Phi_1|^2 + |\Phi_2|^2 = - l^2 \quad\text{and}\quad |Z_1|^2 + |Z_2|^2 + |Z_3|^2 = l^2 \,.
\ee
We will work with global coordinates in $AdS_5$ and this corresponds to the parametrization:
\be
\begin{aligned}
\label{PhiZdefns}
\Phi_0 &= l\, \cosh\rho\, e^{i\phi_0}\qquad \Phi_1 = l\, \sinh\rho\, \cos\theta\, e^{i\phi_1}\qquad \Phi_2 = l\, \sinh\rho\, \sin\theta\, e^{i\phi_2} \,.\cr
Z_1 &= l\, \sin\alpha\, e^{i\xi_1} \qquad\ \  Z_2 = l\, \cos\alpha\, \sin\beta\, e^{i\xi_2}\qquad\ \  Z_3 = l\, \cos\alpha\, \cos\beta\, e^{i\xi_3} \,.
\end{aligned}
\ee
The metric on $AdS_5 \times S^5$ is then simply inherited from the flat metric of the ambient space and takes the following form in global coordinates: 
\begin{multline}
\label{gddbulk}
\frac{ds^2}{l^2} = -\cosh^2\rho \, d\phi_0^2 + d\rho^2+\sinh^2\rho \, (d\theta^2+\cos^2\theta \, d\phi_1^2+\sin^2\theta \, d\phi_2^2)\cr
+ d\alpha^2 + \sin^2\alpha \, d\xi_1^2+\cos^2\alpha \, (d\beta^2 +\sin^2\beta \, d\xi_2^2+\cos^2\beta \, d\xi_3^2)\,,
\end{multline}
where $\phi_0 = \frac{t}{l}$. We choose a frame that makes manifest the fact that $AdS_5$ (respectively $S^5$) can be written as a U(1) Hopf fibration over a K\"ahler manifold $\widetilde{\mathbb{CP}}^2$ (respectively $\mathbb{CP}^2$).  The frame for the $AdS_5$ part is given by 
\be
\begin{aligned}\label{adsframe}
e^0 &= l [\cosh^2\rho \, d \phi_0 - \sinh^2\rho \, (\cos^2\theta d\phi_1 + \sin^2\theta d \phi_2)], \cr 
e^1 &= l \, d\rho\,,\qquad e^2 = l \sinh \rho \, d\theta, \cr
e^3 &= l \cosh\rho \sinh\rho \, ( \cos^2\theta ~ d\phi_{01} + \sin^2\theta  ~ d\phi_{02})\, \cr
e^4 &= l \sinh\rho \, \cos\theta \sin\theta ~ d\phi_{12}
\end{aligned}
\ee
where $\phi_{ij} = \phi_i - \phi_j$.
For the $S^5$ part, we choose the frame
\be
\begin{aligned}\label{sframe}
e^5 &= l \, d\alpha, \qquad e^6 = l \, \cos\alpha \, d\beta, \cr
e^7 &= l \, \cos\alpha \sin\alpha \, (\sin^2\beta \, d\xi_{12} + \cos^2\beta \, d\xi_{13}), \cr
e^8 &= l \, \cos\alpha \cos\beta \sin\beta \, d\xi_{23}, \cr
e^9 &=  l \, (\sin^2\alpha \, d\xi_1 + \cos^2\alpha \sin^2\beta \, d\xi_2 +  \cos^2\alpha \cos^2\beta \,d\xi_3)
\end{aligned}
\ee
where $\xi_{ij} = \xi_i - \xi_j$. 

The Killing spinor for the $AdS_5 \times S^5$ background adapted to the above frame  is given by \cite{Ashok:2008fa}:
\be
\begin{aligned}
\label{adskss2}
\epsilon &=  e^{-\frac{1}{2} (\Gamma_{79} - i \Gamma_5 \, \tilde \gamma) \, \alpha}
e^{-\frac{1}{2} (\Gamma_{89} - i \Gamma_6 \tilde \gamma) \beta} \, e^{\frac{1}{2} 
\xi_1 \Gamma_{57}} \, e^{\frac{1}{2} \xi_2  \Gamma_{68}} \,
e^{\frac{i}{2} \xi_3  \Gamma_9 \, \tilde \gamma} \\
&\hspace{2cm} \times e^{\frac{1}{2} \rho \, (\Gamma_{03} + i \Gamma_1 \,
\gamma)} \,  e^{\frac{1}{2}
\theta \, (\Gamma_{12} + \Gamma_{34})} \, e^{\frac{i}{2} \phi_0 \, \Gamma_0 \, \gamma} \, e^{-\frac{1}{2} \phi_1 \Gamma_{13}} \,
e^{-\frac{1}{2} \phi_2 \Gamma_{24}} \, \epsilon_0 
 \equiv M\cdot \epsilon_0 \,,
\end{aligned}
\ee
where $\epsilon_0$ is an arbitrary 32-component Weyl spinor satisfying $\Gamma_0 \cdots \Gamma_9 \epsilon_0 = - \epsilon_0$ and we have denoted $\gamma = \Gamma^{01234}$ and $\tilde \gamma = \Gamma^{56789}$. 

\subsection{\texorpdfstring{$\frac{1}{2}$}{1by2}-BPS D3-brane Probes}

We now consider various classes of $\frac{1}{2}$-BPS probe D3-branes, all of which end on the boundary in a two dimensional surface. In the first class we consider D3-branes described by the equations:
\be
\label{D3Z1}
\Phi_1\, Z_1 = C_1\,, \quad Z_2=Z_3 = 0\,.
\ee
The embedding equation is inspired by the profile of the complex scalar $Z_1$ in \eqref{class1def}. To be precise, the coefficient that appears in the probe equation and the constant $c_1$ in the profile of the scalar field (see equation \eqref{class1def}) are related by a non-trivial factor, given by \cite{Drukker:2008wr}
\be
\label{mapofparam}
c_1 =\frac{ \sqrt{\lambda}}{2\pi} C_1~,
\ee
where $\lambda = g_{YM}^2N$ is the 't Hooft coupling of the gauge theory. The relative factor in the normalization can be explained as follows: the probe D3 brane action naturally comes with an overall factor that is the tension of the D3 brane given by  $T_{D3} = \frac{N}{2\pi^2 \, l^4}$. If we consider the boundary limit of the probe brane action, and require the action to reproduce the action for a single eigenvalue of the complex scalar field $Z$ in the boundary theory, then, this is precisely the factor one would rescale the field by in order to obtain the action\footnote{Here, we identify the radial profile of the brane probe in $AdS$ is identified with $|Z|$ on the CFT side. } in \eqref{cft-lagrangian}. Alternatively, we could simply declare this to be the map between the classical solutions of the probe theory and in the boundary theory, in which a single eigenvalue is given a non-trivial profile. In either case, we shall see that this map of parameters is essential to match the energies and charges computed in the bulk and boundary theories, in the leading order expansion in $\lambda$.  

In terms of the real coordinates introduced in the previous subsection, the embedding of the probe D3-brane is given by the following real conditions:
\begin{align}
\label{D3Z1real}
\sinh\rho\, \cos \theta = \frac{R_0}{l}\,, \quad  \alpha = \frac{\pi}{2}\,,  \quad  \phi_1 + \xi_1 = \xi_1^{(0)} \,.
\end{align}
We have chosen to write the complex constant $C_1=R_0\, e^{i\xi_1^{(0)}}$ in a particularly convenient manner. We see that as $\rho\rightarrow \infty$, we have $\theta \rightarrow \frac{\pi}{2}$ so as to keep the first equation consistent, and the D3-brane ends on the circle parametrized by $\phi_1$ on the boundary, while being extended along the $\phi_0$-direction. The $(\theta, \phi_2)$ coordinates parametrize the directions transverse to the boundary limit of the probe, exactly as for the corresponding string defect. We choose the static gauge in which the world-volume coordinates are identified as follows: 
\be
\label{staticgauge}
(\tau, \sigma_1, \sigma_2, \sigma_3) = (\phi_0, \theta, \phi_1, \phi_2) \,.
\ee
The induced metric on the world-volume is given by: 
\be
\label{hinducedclass1}
\left. \frac{ds^2}{l^2}\right|_{D3} = -\left( \frac{R_0^2+l^2\cos^2\theta}{l^2\cos^2\theta} \right)d\phi_0^2 + \frac{R_0^2(R_0^2+l^2)\sec^2\theta}{l^2(R_0^2+l^2\cos^2\theta)}d\theta^2 + \frac{R_0^2+l^2}{l^2}d\phi_1^2 + \frac{R_0^2}{l^2}\tan^2\theta d\phi_2^2\,.
\ee
The square root of the determinant of the induced metric, which will play an important role, takes the simple form 
\be
\label{dethcalc}
\sqrt{-\text{det}(h)} = R_0^2(R_0^2+l^2)\sec^2\theta\tan\theta \,. 
\ee

\subsubsection{The $\kappa$-symmetry analysis}

We would now like to classify the set of supersymmetries preserved by this probe D3-brane. The $\kappa$-symmetry equation that guarantees the supersymmetry of the worldvolume theory is given by 
\begin{align}
\label{kappaconstraint}
\gamma_{\tau\sigma_1 \sigma_2 \sigma_3} \epsilon =  \pm \, i \, \sqrt{- \det h} \,\, \epsilon~.
\end{align}
Here, the world-volume $\gamma$-matrices are defined by
\be
\gamma_{i} = \mathfrak{e}^a_i\, \Gamma_a\,,
\ee
where the $\mathfrak{e}^a_i = e^a_{\mu}\pa_i X^{\mu}$ is obtained by the pullback of the one-form $e^{a}_{\mu}$. For the probe D3-brane under consideration, the world-volume gamma matrices are as follows: 
\be
\begin{aligned}
  \gamma_{\tau} &=  l\,\cosh^2\rho\,\Gamma_0 +l \sinh\rho \cosh\rho \, \Gamma_3~,  \cr
  \gamma_{\sigma_1} &= l \tanh\rho \, \tan \theta \, \Gamma_1 +l \sinh\rho  \, \Gamma_2~,  \cr
  \gamma_{\sigma_2} &= - l\, \sinh^2\rho \cos^2 \theta \, \Gamma_0 - l \sinh\rho \cosh\rho \cos^2 \theta \, \Gamma_3  +  l \sinh\rho \cos \theta \sin \theta \, \Gamma_4 
   - \, l \, \Gamma_9 ~,\cr
  \gamma_{\sigma_3} &= - l \sinh^2\rho \sin^2 \theta \, \Gamma_0 - l\sinh\rho\cosh\rho \sin^2 \theta \, \Gamma_3  -  l\sinh\rho \cos \theta \sin \theta \, \Gamma_4~.
\end{aligned}
\ee
The product of four $\gamma$ matrices is
\be
\begin{aligned}
  \label{gammats123}
\frac{1}{l^4} \gamma_{\tau \sigma_1 \sigma_2 \sigma_3} &=  \sinh^2\rho \cosh\rho \bigg[ \big(\sinh\rho( \Gamma_{0234} +\Gamma_{2349})- \,  \cosh\rho \, \Gamma_{0249} \big)
  \cos \theta \sin \theta \,- \Gamma_{0239} \, \sin^2 \theta \bigg] \cr
   & \quad + \sinh^2\rho \sin^2 \theta \bigg[ \sinh\rho\, (\Gamma_{0134}+ \Gamma_{1349}) - \cosh\rho \, \Gamma_{0149}  -\tan \theta \, \Gamma_{0139}\bigg]  \,.
\end{aligned}
\ee
 In order to check the $\kappa$-symmetry equation, we need to commute the four-gamma products through the matrix $M$ defined in \eqref{adskss2}. For instance, we have the following identity: 
 \be
 \begin{aligned}
 & \Gamma_9 \, M = M \, \bigg( \mu_3^2 \,-\, i \mu_2^2 \, \Gamma_{57} \,-\, i \, \mu_1^2 \, \Gamma_{68}  \,+\, \mu_2 \mu_3 e^{-\xi_2\,\Gamma_{68} - i\xi_3\Gamma_{5678}} 
  \left( 1 + i \Gamma_{57}  \right) \Gamma_{89} \cr
 & \hspace{1.5cm} \,+\, \mu_1 \mu_3 e^{-\xi_1\,\Gamma_{57} - i\xi_3\Gamma_{5678}} \left( 1 + i \Gamma_{68} \right) \Gamma_{79} 
 \,+\, i \mu_1 \mu_2 e^{-\xi_1\,\Gamma_{57} -\xi_2\,\Gamma_{68}} \left( \Gamma_{58} + \Gamma_{67} \right)  \bigg)\, \Gamma_9 \,. \cr
 \end{aligned}
 \ee
 Here we have defined $\mu_1= \sin\alpha$, $\mu_2= \cos\alpha\sin\beta$ and $\mu_3 = \cos\alpha\cos\beta$ so that $Z_i = \mu_i\, e^{i\, \xi_i}$ in \eqref{PhiZdefns}. For the particular probe brane under consideration, we have $\alpha=\frac{\pi}2$, and this corresponds to $\mu_2 = \mu_3=0$ and $\mu_1=1$. This simplifies the above relation to 
 \begin{align}
  \Gamma_9 \, M &= - i\, M \, \Gamma_{689}\,.
 \end{align}
 One can similarly commute the other $\Gamma$-matrices through the matrix $M$. After some tedious algebra, the $\kappa$-symmetry constraint reduces to the following simple expression: 
 \be
 \begin{aligned}
 \frac{1}{l^4} \gamma_{\tau \sigma_1 \sigma_2 \sigma_3}\cdot M\cdot \epsilon_0 =& M\, \cosh\rho\sinh^3\rho\, e^{-i\phi_0 \Gamma_0\gamma}\, e^{\phi_1\Gamma_{13}}(\Gamma_{0234}+\Gamma_{3968})\cdot \epsilon_0 \cr
 &- i M\, \sin^2\theta\sinh^4\rho\, e^{\phi_1\Gamma_{13}}\, e^{\phi_2\Gamma_{24}}(\Gamma_{12}+\Gamma_{014968}) \cdot\epsilon_0\cr
&+i M\, \tan\theta \sinh^2\rho(1+\cos^2\theta \sinh^2\rho) \Gamma_{024968} \cdot\epsilon_0~.
 \end{aligned}
 \ee
 Using the embedding equation in \eqref{D3Z1real} and the 10d chirality constraint, we find that the $\kappa$-symmetry constraint in \eqref{kappaconstraint} is satisfied with the choice of $(-)$ sign if the following projection constraint is imposed on the constant spinor $\epsilon_0$: 
 \be
 \Gamma_{1357} \, \epsilon_0 = \epsilon_0 \,.
 \ee
 We have thus shown that the probe D3-brane preserves half of the bulk supersymmetries.  
 
  \subsubsection{More $\frac{1}{2}$-BPS Probes from SU$(3)$ Rotations}

The advantage of the coordinates and frame we have chosen to work with is that it is possible to find other probe D3-branes that are closely related to the one we have analyzed so far, and whose supersymmetry can be checked by a minor modification of our previous analysis. These probes are obtained by using an SU$(3)$ rotation acting on the $Z_i$ variables and as a result the induced metric remains the same as in \eqref{hinducedclass1}. 

Repeating the $\kappa$-symmetry analysis we find that 
\begin{align}
\Phi_1\, Z_2 &= C_2\quad\text{and}\quad Z_1=Z_3=0 
\end{align}
is half-BPS and preserves the supersymmetries that survive the following projection:
 \be
 \Gamma_{1368}\epsilon_0 = \epsilon_0 \,.
 \ee
 Similarly the probe D3-brane 
\begin{align}
\Phi_1\, Z_3 &= C_3\quad\text{and}\quad Z_1=Z_2=0\,.
\end{align}
preserves half the supersymmetries if we impose the projection
 \be
 \Gamma_{0924}\epsilon_0 = i\, \epsilon_0 \,.
 \ee

\subsubsection{A Second Class of $\frac{1}{2}$-BPS D3-branes} 
  
We now mirror our discussion of classical string like defects on the boundary theory and turn to discuss a second class of D3 probe branes, that are obtained by an SU$(2)$ rotation that acts on the complex $\Phi_i$ variables. While the $\kappa$-symmetry analysis is similar to the one performed earlier, the technical details are quite different. Consider now the following probe D3-brane:
\be
\Phi_2 Z_1 = D_1\quad\text{and}\quad Z_2=Z_3=0\,.
\ee
In terms of the real coordinates we now have the defining equations:
\be
\sinh\rho\sin\theta = \frac{R_0}{l} \qquad \alpha=\frac{\pi}{2}\qquad \phi_2+\xi_1 = \xi^{(0)} \,.
\ee
The induced metric on the worldvolume is given by
\be
\label{hinducedclass2}
\left. \frac{ds^2}{l^2}\right|_{D3} = -\left( \frac{R_0^2+l^2\sin^2\theta}{l^2\sin^2\theta} \right)d\phi_0^2 + \frac{R_0^2(R_0^2+l^2)\csc^2\theta}{l^2(R_0^2+l^2\sin^2\theta)}d\theta^2 + \frac{R_0^2+l^2}{l^2}d\phi_1^2 + \frac{R_0^2}{l^2}\cot^2\theta d\phi_2^2\,.
\ee
The square root of the determinant of the induced metric is given by: 
\be
\sqrt{-\text{det}(h)} = R_0^2(R_0^2+l^2)\csc^2\theta\cot\theta \,. 
\ee
The world-volume gamma matrices are as follows: 
\be
\begin{aligned}
 \gamma_{\tau} &=  l\,\cosh^2\rho\,\Gamma_0 +l \sinh\rho \cosh\rho \Gamma_3  \cr
  \gamma_{\sigma_1} &=- l \tanh\rho \, \cot \theta \, \Gamma_1 +l \sinh\rho  \, \Gamma_2  \cr
  \gamma_{\sigma_2} &= - l\, \sinh^2\rho \cos^2 \theta \, \Gamma_0 - l \sinh\rho \cosh\rho \cos^2 \theta \, \Gamma_3  +  l \sinh\rho \cos \theta \sin \theta \, \Gamma_4 
   \cr
  \gamma_{\sigma_3} &= - l \sinh^2\rho \sin^2 \theta \, \Gamma_0 - l\sinh\rho\cosh\rho \sin^2 \theta \, \Gamma_3  -  l\sinh\rho \cos \theta \sin \theta \, \Gamma_4 - \, l \, \Gamma_9
\end{aligned}
\ee
The product of four $\gamma$ matrices is
\begin{align}
%  \label{gammats123}
\frac{1}{l^4} \gamma_{\tau \sigma_1 \sigma_2 \sigma_3} =& -\cos^2\theta\sinh^2\rho\left(\cot\theta \Gamma_{0139}- \cosh\rho(\Gamma_{0149}+\Gamma_{0239})+\sinh\rho(\Gamma_{0134}+\Gamma_{1349}) \right) \cr
&-\sin\theta\cos\theta\cosh\rho\sinh^2\rho(\cosh\rho\Gamma_{0249}-\sinh\rho(\Gamma_{0234}+\Gamma_{2349}) ) \,.
\end{align}
In order to check the $\kappa$-symmetry equation, as before, we need to commute the four-gamma products through the matrix $M$ defined in \eqref{adskss2}. After performing the relevant $\Gamma$-matrix algebra, we finally obtain 
\be
 \begin{aligned}
 \frac{1}{l^4} \gamma_{\tau \sigma_1 \sigma_2 \sigma_3}\cdot M\cdot \epsilon_0 =& i\, M\sinh^3\rho\cos\theta\left[ i\cosh\rho e^{-i \phi_0\Gamma_0\gamma}e^{\phi_2\Gamma_{24}}(\Gamma_{0134}-\Gamma_{4968}))  \right. 
\cr
& \left. +\sinh\rho\cos\theta e^{\phi_1\Gamma_{13}}\, e^{\phi_2 \Gamma_{24}}(\Gamma_{12} - \Gamma_{023968}) \right]\cdot\epsilon_0
\cr
&+i M\, \cot\theta \sinh^2\rho(1+\sin^2\theta \sinh^2\rho) \Gamma_{013968} \cdot\epsilon_0
 \end{aligned}
 \ee
We thus find that the D3-brane preserves one half of the bulk supersymmetries if the following projection is imposed on the constant spinor:
 \begin{align}
 \Gamma_{2457} \, \epsilon_0 &=\epsilon_0 \,.
 \end{align}

\subsubsection{More $\frac{1}{2}$-BPS Probes from SU$(3)$ Rotations}
 
One can now do an SU$(3)$ rotation on the $Z_i$ variables as before and obtain two other half-BPS probe D3-branes in this same class. These also turn out to be half-BPS; we find that 
 \begin{align}
 \Phi_2 Z_2 &=D_2\quad\text{and}\quad Z_3=Z_1=0\,,
 \end{align}
 preserves half the supersymmetries if the following projection is imposed on the constant spinor: 
 \begin{align}
 \Gamma_{2468} \, \epsilon_0 &=\epsilon_0 \,.
 \end{align}
 Similarly, the D3-brane described by
\be
\Phi_2 Z_3 =D_3\quad\text{and}\quad Z_1=Z_2=0\,,
\ee
preserves half the supersymmetries if the following projection is imposed on the constant spinor:
\begin{align}
\Gamma_{0913} \, \epsilon_0 &= i \, \epsilon_0\,. 
 \end{align}

 \subsection{\texorpdfstring{$\frac{1}{16}$}{1by16}-BPS D3-brane Probes}
 \label{generalBPSworldvolume}
 
So far we have found  six different probe D3-branes, that have been classified into two distinct classes depending on whether it wraps the $\phi_1$ circle or the $\phi_2$ circle on the boundary. Each probe brane preserves half of the supersymmetries. Following our analysis of the singular solutions on the boundary, we now ask for the projections that preserve the common set of supersymmetries amongst all these probe D3-branes. This is easily done and the result is the following set of projections on the constant spinor:
\be
\label{generalprojection}
\Gamma_{13}\epsilon_0 = \Gamma_{24}\epsilon_0 = -i\epsilon_0\,, \qquad \Gamma_{09}\epsilon_0 = -\epsilon_0\,, \qquad \Gamma_{57}\epsilon_0 = \Gamma_{68}\epsilon_0 = i \epsilon_0 \,.
\ee
These projections preserve exactly two out of the thirty two supersymmetries of the bulk background. 

Remarkably, these projections conditions have been encountered previously in the context of studying giant-gravitons and dual giant-gravitons in $AdS_5 \times S^5$ \cite{Ashok:2008fa}. What we have just shown is that the set of two supersymmetries that the various probe branes share (and which are dual to stringy defects in the gauge theory), is the same set of supersymmetries shared by the D3-brane probes that describe giants and dual-giants. 

We now review the constraints on the D3-brane worldvolume that were derived in \cite{Ashok:2008fa} from the projections in \eqref{generalprojection}. As we shall see, the general solution to these can be suitably described as zeros of holomorphic functions and this in turn will enable us to describe the most general D3 probe that ends on a string like defect on the boundary. 

The projection conditions in \eqref{generalprojection} leads to a drastic simplification of the Killing spinor, which now takes the form:
\be
\label{KSnew}
\epsilon = e^{\frac{i}{2}(\phi_0+\phi_1+\phi_2 + \xi_1+\xi_2+\xi_3)}\, \epsilon_0\,. 
\ee
Our goal is to write down the general conditions satisfied by the D3 world-volume that preserves these two supercharges. The world-volume $\gamma$-matrices are given by
\be
\gamma_{i} = \mathfrak{e}^a_i\Gamma_a\,,
\ee
where $\mathfrak{e}^a_i = e^a_{\mu}\partial_i X^{\mu}$ is the pullback of the spacetime frame $e^a_{\mu}$ onto the world-volume. The $\kappa$-projection condition is given by
\be
\label{kappasymmgeneral}
\gamma_{\tau\sigma_1\sigma_2\sigma_3}\ \epsilon = \pm i \sqrt{-\det h}\ \epsilon \,.
\ee
Using the definitions, we substitute the Killing spinor \eqref{KSnew} into \eqref{kappasymmgeneral} and use the projection conditions to reduce the l.h.s. into a linear combination of independent structures of the form $\Gamma_{a_1, a_2\ldots} \epsilon_0$. The coefficient of each such structure is set to zero except the constant one, which is equated to the r.h.s. 

In order to write these BPS equations in a compact form, we introduce the following complex $1$-forms:
\begin{equation}\label{cplxforms}
{\bf E}^1 = {\mathfrak e}^1-i {\mathfrak e}^3\qquad {\bf E}^2 = {\mathfrak e}^2-i {\mathfrak e}^4\qquad {\bf E}^5= {\mathfrak e}^5+i {\mathfrak e}^7\qquad {\bf E}^6= {\mathfrak e}^6+i {\mathfrak e}^8 \,,
\end{equation}
We then find that the $\kappa$-symmmetry constraints that follow by setting to zero the coefficient of $\Gamma_{a_1, \ldots a_n}\epsilon_0$ are equivalent to the vanishing of the pullback of the following 4-forms onto the D3 world-volume \cite{Ashok:2008fa}: 
\begin{align}
\label{compactbps}
{\bf E}^{ABCD}&= 0 \cr
({\mathfrak e}^{09}+i\, (\tilde{\bf \omega}- {\bf \omega})) \wedge {\bf E}^{AB} &= 0 \quad \hbox{for} \quad A, B = 0, 1, 2, 5, 6\, .
\end{align}
Here we have also defined the following real 2-forms: 
\begin{align}
\tilde{\bf \omega} &= {\mathfrak e}^{13}+ {\mathfrak e}^{24}= -\frac{i}{2}\left({\bf E}^1\wedge \overline{{\bf E}^1} + {\bf E}^2\wedge \overline{{\bf E}^2}\right) \equiv \omega_{\widetilde{\mathbb{CP}}^2}\\
{\bf \omega}&= {\mathfrak e}^{57}+ {\mathfrak e}^{68}=\frac{i}{2}\left({\bf E}^5\wedge \overline{{\bf E}^5} + {\bf E}^6\wedge \overline{{\bf E}^6}\right) \equiv \omega_{\mathbb{CP}^2}\,.
\end{align}
As the notation suggests, the $2$-forms are the pull-backs of certain K\"ahler forms onto the worldvolume of the brane. These K\"ahler forms are of the respective base manifolds $\mathbb{CP}^2$ and $\widetilde{\mathbb{CP}}^2$, when $S^5$ and $AdS_5$ are written as Hopf-fibrations. 

Substituting the equations in \eqref{compactbps} into the $\kappa$-symmetry constraint and equating the coefficient of $\epsilon_0$ on both sides, we find that for D3 probes that have a time-like world-volume we have
\begin{align}
\label{timelikebps}
({\bf \omega} - \tilde{\bf \omega})\wedge ({\bf \omega}-\tilde{\bf \omega}) &= 0\\
{\mathfrak e}^{09} \wedge (\tilde{\bf \omega} - {\bf \omega}) = \pm \left\vert {\mathfrak e}^{09} \wedge (\tilde{\bf \omega}-{\bf \omega}) \right\vert &= \pm \text{dvol}_4~.
\label{volumeform}
\end{align}
The equation \eqref{volumeform} is solved for either D-branes or anti D-branes depending on the sign of  $\vert {\mathfrak e}^{09} \wedge (\tilde{\bf \omega} -{\bf \omega})\vert$. For dual-giant solutions, one can check that it is for the negative sign ({\it i.e.} for anti-branes) that the $\kappa$-symmetry conditions are satisfied. 

This completes the identification of the volume element on the world volume of the D3-brane as the pullback of a particular spacetime 4-form. This will have important consequences when we discuss the on-shell actions for our probe brane solutions. We have been brief in this review of the BPS equations and refer the reader to \cite{Ashok:2008fa} for the complete derivation. 

\subsubsection{General $\frac{1}{16}$-BPS Solutions}

The most general $\frac{1}{16}$-BPS solution to these equations were given by Kim and Lee \cite{Kim:2006he} (see also  \cite{Ashok:2008fa}) in terms of three holomorphic functions: 
\be
F^{(I)}(\Phi_i, Z_j) = 0\quad\text{for}\quad I =1,2,3 \,,
\ee
where the $\Phi_i$ and $Z_j$ are defined in \eqref{PhiZdefns} and the functions each satisfy a scaling condition:
\be
\sum_{i=0}^2 \pa_{\phi_i} F^{(I)} -\sum_{i=1}^3  \pa_{\xi_i} F^{(I)} = 0 \,.
\ee
Four sub-classes of solutions to these equations that preserve $\frac{1}{8}$th of the bulk supersymmetry were listed in \cite{Ashok:2008fa}, some of which were previously obtained in \cite{Mikhailov:2000ya, Arapoglu:2003ti, Caldarelli:2004yk, Mandal:2006tk}. The probes were either point-like in the $AdS_5$ directions (giants) or point-like in the $S^5$ (dual-giants) and they carried  spins $(J_1, J_2, J_3)$ only along the $S^5$ or they carried two spins along the $AdS_5$ directions and one spin along the $S^5$, which we denote $(S_1, S_2, J)$. 

All the particular solutions that were considered in \cite{Ashok:2008fa} had compact world-volume and none of these extended to the boundary. What we have just shown is that the same set of BPS equations admit another completely different class of probe D3-branes that have an interpretation as holographic duals of string like defects, and  whose world-volume ends on the conformal boundary $\mathbb{R}\times S^3$ along two directions, one of which is the time direction. We next turn to a better understanding of the constraints that holomorphy places on the spatial direction of the boundary component of the probe brane.

\subsection{Bulk Zeros to Boundary Profiles}
\label{holostrings}

The particular noncompact solutions of the probe D3-branes we are interested in and which are the holographic duals of the wobbling stringy solutions we found on the boundary have charges $(S_1, S_2, J)$. From the above discussion, these should therefore be described by
\be
\label{wobstr}
Z_2=Z_3=0\quad\text{and}\quad f(Z_1\Phi_0 , Z_1\Phi_1, Z_1\Phi_2) = 0 ~.
\ee 
We have written the holomorphic function in such a way that it is invariant under the scaling $$\Phi_i \rightarrow \lambda \, \Phi_i ~~~ {\rm and} ~~~ Z_1 \rightarrow \lambda^{-1} Z_1~.$$ 
Here and in what follows, we shall omit the subscript and simply refer to the coordinate on the $S^5$ as $Z$. 
We could also rewrite the holomorphic function in \eqref{wobstr} in a manner suggestive of the boundary solutions, as functions of the form $g( \Phi_0 Z, \Phi_1/\Phi_0, \Phi_2/\Phi_0)$.

Our goal is to find the zeros of this function on the boundary of $AdS_5$. What we shall then show is that this zero locus precisely coincides with the locus in the boundary theory where the profile of the scalar field has a singularity. For this purpose let us parametrise the coordinates $\Phi_i \in {\mathbb C}^{1,2}$ of the ambient space as follows:
\bea
\Phi_0 = \sqrt{r^2 +l^2} ~ \nu_0, ~~ \Phi_1 = r \, \nu_1, ~~ \Phi_2 = r \, \nu_2 \, 
\eea
where $\nu_0 =e^{i\phi_0}$, $\nu_1 = \cos\theta \, e^{i\phi_1}$ and $\nu_2 = \sin\theta \, e^{i\phi_2}$ so that we have $-|\Phi_0|^2 + |\Phi_1|^2 + |\Phi_2|^2 = -l^2$. As we near the boundary these take the form
\bea
\Phi_0 = r\,  \nu_0, ~~ \Phi_1 = r \, \nu_1, ~~ \Phi_2 = r \, \nu_2 \,, 
\eea
and become coordinates on a null-cone $-|\Phi_0|^2 + |\Phi_1|^2 + |\Phi_2|^2 = 0$. The induced metric on this cone is of the form
\be
\begin{aligned}
-|d\Phi_0|^2+ |d\Phi_1|^2+ |d\Phi_2|^2 &= r^2 \, (- |d\nu_0^2| + |d\nu_1|^2 + |d\nu_2|^2) \\
&= r^2 \, (-d\phi_0^2 + d\theta^2 + \cos^2\theta \, d\phi_1^2 + \sin^2\theta \, d\phi_2^2)~,
\end{aligned}
\ee
so that the boundary is in the conformal class of $\mathbb{R}\times S^3$ for {\it arbitrary} and large-$r$. 

Now that we have obtained the asymptotic behaviour of the bulk coordinates let us return to the problem at hand, which is to find the locus of zeros of $f(Z \Phi_0, Z \Phi_1,Z \Phi_2)$ as we approach the boundary. Near the boundary, the function becomes $f(Z \, r \, \nu_0, Z \, r \nu_1,Z \, r \, \nu_2)$ with $Z = e^{i\xi}$. So the worldvolume of the D3 brane intersects the boundary at the zeros of the functions 
$$
f(\lambda \, \nu_0, \lambda \, \nu_1, \lambda \, \nu_2) =0~,
$$
where $\lambda = r \, e^{i \, \xi}$ for arbitrary $\lambda \in {\mathbb C}^\star$. Such a zero set remains invariant only if this function is homogeneous under scaling, which means $f(\lambda \, \nu_0, \lambda \, \nu_1, \lambda \, \nu_2) = \lambda^p f (\nu_0, \nu_1,  \nu_2)$. But a function with such a scaling property can be re-written as $\nu_0^p\, F(\nu_1/\nu_0, \nu_2/\nu_0)$ so that the zeros we are after at a fixed $\tau$ is also the same as the zeros of a holomorphic function $F(\zeta_1, \zeta_2)$ where $\zeta_i  = \nu_i/\nu_0 \in {\mathbb C}^2$ which intersects the unit 3-sphere $|\zeta_1|^2 + |\zeta_2|^2 =1$. The time evolution is simply given by the scaling $(\zeta_1, \zeta_2) \rightarrow e^{-i \phi_0} (\zeta_1, \zeta_2)$.

Let us summarize what we have just derived. If the worldvoume of the probe D3-brane is described as the zero locus of an arbitrary holomorphic function $f(Z \Phi_0, Z \Phi_1,Z \Phi_2)$, then we see that for those probes that reach the boundary, the world-volume, as it approaches the boundary is two dimensional and at a given instant in time, it is given by the locus ${\cal K}$, which, following the steps outlined above, is obtained by the intersection of a holomorphic function in $\mathbb{C}^2$ with the 3-sphere.
\be
F(\zeta_1, \zeta_2) = 0\quad \cap\quad |\zeta_1|^2  + |\zeta_2|^2= 1 ~.
\ee
The curve ${\cal K}$ is an algebraic link in $S^3$ (see for instance \cite{Milnor}). We recall that we found a very similar characterization in the analysis of the boundary theory  in Section \ref{giantdefects}, in which precisely this one dimensional locus was found as the spatial part of the worldvolume of the wobbling string solution. This was the locus where the profile of the complex scalar field $Z$ of the ${\mathcal N}=4$ theory became singular. So, to complete our bulk analysis what we need to do is to derive this boundary profile from the zeros of the holomorphic function.  

Let us start with the bulk solution but now rewrite it in the following manner:
$$
f(Z \Phi_0, Z \Phi_1,Z \Phi_2) =g(Z \Phi_0, \Phi_1/\Phi_0, \Phi_2/\Phi_0)=0~.
$$ 
Since this function $g$ is considered to be a polynomial in the variable $Z\Phi_0$ of degree, say, $p\le N$, it can be factorised as 
\be
\label{gfactored}
g(Z \Phi_0, \Phi_1/\Phi_0, \Phi_2/\Phi_0) = \prod_{r=1}^p \Big[ (Z \Phi_0) \, F^{(1)}_r ( \Phi_1/\Phi_0, \Phi_2/\Phi_0)- F_r^{(0)} (\Phi_1/\Phi_0, \Phi_2/\Phi_0) \Big]~.
\ee
From the discussion of the boundary limits of the coordinates $\Phi_i$, we infer that near the boundary, this function becomes 
\be
g(Z \Phi_0, \Phi_1/\Phi_0, \Phi_2/\Phi_0) \longrightarrow  (\lambda \nu_0)^p \prod_{r=1}^p  \, F^{(1)}_r ( \nu_1/\nu_0, \nu_2/\nu_0)~.
\ee
Here $\lambda = r\, e^{i\xi}$ is a field on the probe brane that determines the radial and angular profile of the probe brane and we will identify it with the complex scalar field denoted by $Z$ in the boundary theory. 
The defects on the boundary are therefore given by zero-sets of $F^{(1)}_r (\nu_1/\nu_0, \nu_2/\nu_0)$. So far we reproduced the conclusion of the bulk analysis.  

As a first step towards deriving the boundary profile, let us set $p=1$ for simplicity.
Then the bulk solution $g(Z \Phi_0, \Phi_1/\Phi_0, \Phi_2/\Phi_0)$ is of unit degree in $Z$: 
\be
g(Z \Phi_0, \Phi_1/\Phi_0, \Phi_2/\Phi_0) = Z\Phi_0 \, F_1(\Phi_1/\Phi_0, \Phi_2/\Phi_0) - F_0(\Phi_1/\Phi_0, \Phi_2/\Phi_0) =0
\ee
and very near (but not exactly at) the boundary this is equivalent to
\be
\lambda \, \nu_0 = \frac{F_0(\nu_1/\nu_0, \nu_2/\nu_0)}{F_1(\nu_1/\nu_0, \nu_2/\nu_0)}~.
\ee
From a single probe brane one therefore infers the boundary profile that corresponds to one of the eigenvalues of the scalar field $Z$ of the boundary theory. For degree $p>1$ and for generic polynomials, it follows that the resulting holomorphic function can be factorized, as in \eqref{gfactored}, and each of the linear factors lead to profiles for $p$ of the eigenvalues of the matrix valued field $Z$. Given that $Z$ is an $N\times N$ matrix, this leads to $p\le N$ and is referred to as the stringy exclusion principle  \cite{Bena:2004qv, Suryanarayana:2004ig, Mandal:2006tk, Ashok:2008fa}. 

\section{Holographic Wobbling Strings}
\label{wobblers}

We now turn to the holographic description of the monomial type defect solutions in the ${\mathcal N}=4$ gauge theory and compute the holographically renormalized energies from the probe D3-brane point of view.  
We find it convenient to redefine the radial coordinate $r = l\, \sinh\rho\,,$ and work with the following metric on $AdS_5\times S^5$:
\be
ds^2_{AdS_5} = -V(r) \, dt^2 + \frac{dr^2}{V(r)} +r^2 (d\theta^2 + \cos^2 \theta \, d\phi_1^2 + \sin^2\theta \, d\phi_2^2)~,
\ee
where $V(r)=1+ \frac{r^2}{l^2}$. The Ramond-Ramond 4-form in these coordinates is given by
\be
C_{(4)} = -\frac{r^4}{l} \cos\theta\sin\theta\, dt\wedge d\theta\wedge d\phi_1\wedge d\phi_2~.
\ee
The Lagrangian density for a probe D3-brane is:
\bea
{\cal L} = -T_{D3}\, \sqrt{- h} + T_{D3}\, P[C^{(4)}]
\eea
where $h$ is the determinant of the induced metric on the worldvolume and $P[\cdot]$ refers to the pullback of a spacetime differential form onto the worldvolume. In order to make the formulae less cumbersome, we shall omit the factor $T_{D3} = \frac{N}{2\pi^2 \, l^4}$ for now and restore it later on when the energies and charges are evaluated.

The probe D3-branes we are interested in are given by the following monomial type solution:
\be
\label{probemono}
(Z_1 \Phi_0) =  \eta\, \left(\frac{\Phi_1}{\Phi_0}\right)^m \, \left(\frac{\Phi_2}{\Phi_0}\right)^n~, \quad\text{and}\quad~ Z_2=Z_3=0~.
\ee
This follows from our general analysis in the previous section and the classical profile in \eqref{solution1}. For such a probe brane described by \eqref{probemono} we choose the worldvolume coordinates to be $(\phi_0=\frac{t}{l}, \theta,\phi_1, \phi_2)$, with $r=r(\theta)$ and $\xi = \xi(\phi_0,\phi_1,\phi_2)$ as the fluctuating fields on the worldvolume. 

If we set the parameter $\eta = R_0 \, e^{i \xi_0}$, then the defining equation of the probe D3-brane can be written as the pair of real equations:
\be
\begin{aligned}
\label{rdef}
\left(\frac{l}{r}\right)^{m+n} \, \left(1+ \frac{r^2}{l^2} \right)^{\frac{1}{2}(m+n+1)}  &= \frac{R_0}{l} \, \cos^m \theta \, \sin^n \theta, \\
\xi- \xi_0  &= m \, \phi_1 + n \, \phi_2 - (m+n+1) \, \phi_0 ~.
\end{aligned}
\ee
By taking derivatives, it is possible to write down the first derivatives as follows:
\bea
\label{bpseqns}
&& \partial_{\phi_1} \xi = m, ~~ \partial_{\phi_2} \xi = n, ~~ \partial_{\phi_0} \xi = - (m+n+1), \cr
&& \cr
&& \partial_\theta r = r \, \left(1+ \frac{r^2}{l^2} \right)  ~\frac{(m \, \sec^2\theta - n \, \csc^2\theta)}{\left(m+n - \frac{r^2}{l^2} \right)} \, \cos\theta \, \sin\theta~.
\eea
It is not possible to solve for $r(\theta)$ in closed form except for a few cases. We will mostly focus on the following three cases:
\begin{enumerate}

\item The simplest case is the static case for which $m+n+1=0$ and we shall consider $n<0$. For this case we have
\be
 r(\theta) = R_0  \, \sec \theta \, \cot^{|n|}\theta~.
 \ee
\item The next simplest case that we shall deal with corresponds to $m=0$ and $n<0$. For this case, one can check that $r(\theta)$ can be expanded order by order in $\frac{l}{R_0}$ as follows:
\be
\label{rformnzero}
r(\theta) = R_0\sin^{-|n|}\theta\left(1-\frac{(n+1)l^2}{2R_0^2} \sin^{2|n|}\theta - \frac{(n+1)(3n+1)l^4}{8R_0^4}\sin^{4|n|}\theta + \ldots \right)
\ee
 \item Lastly we consider those cases for which $m+n =0$ with $n<0$. For this case it is possible to solve for $r(\theta)$ exactly using the defining equation in \eqref{rdef} and we have
 \be
 \label{rmnsumzero}
 r(\theta) = \sqrt{R_0^2 \, \cot^{2|n|} \theta - l^2} ~.
 \ee
\end{enumerate}

\subsection{On-shell Action and the Variational Problem}

Our first task is to define a consistent variational problem such that all the monomial solutions are included in the set of solutions. Given the noncompact nature of the D3-branes, this involves adding appropriate boundary terms near the boundary of $AdS_5$ (equivalently near $\theta=0$ or $\theta=\frac{\pi}{2}$). A simple way to check the consistency of the proposal would then be to ensure that the on-shell action (including both bulk and boundary contributions) for all solutions gives the same value, independent of $(m,n)$ and $\eta = R_0 \, e^{i \xi_0}$. 

From the general analysis of the world-volume action in Section \ref{generalBPSworldvolume} we have seen that the BPS equations simplify the on-shell Lagrangian to take the following form (see equation \eqref{volumeform} for the volume form on the D3-brane):
\be
\label{d3LDensity}
{\cal L} \big|_{\text{on-shell}}= - P\left[e^{09} \wedge (e^{13}+e^{24})\right] + P\left[C^{(4)}\right]~.
\ee
Let us evaluate this for the monomial solution. We work with the ansatz $\xi = \xi (\phi_0, \phi_1, \phi_2)$ and $r = r(\theta)$ suitable for the monomial solution and denote the conjugate momenta as follows:
\be
\Pi^{\mu}_{r} = \frac{\partial {\cal L}}{\partial (\partial_{\mu} r)}~, \qquad \Pi^{\mu}_{\xi} = \frac{\partial {\cal L}}{\partial (\partial_{\mu} \xi)}~.
\ee
Evaluating these on the particular monomial solutions labelled by $(m,n)$, we find the following results:
\be
\begin{aligned}
\label{Pisinbulk}
\Pi^{\tau}_{\xi} &= \frac{(m+n+1) \, r^4\sin2\theta}{2(m+n-\frac{r^2}{l^2})}~,\qquad\Pi^{\theta}_r = \frac{rl^2}{2}\big(m \, \sin^2\theta \, - n \, \cos^2 \theta \big) \\
\Pi^{\phi_1}_{\xi} & = \frac{m\, r^2 \tan\theta \, (r^2+l^2)}{(m+n-\frac{r^2}{l^2})} ~, \qquad \Pi^{\phi_2}_{\xi} = \frac{n\, r^2\, \cot\theta \, (r^2+l^2)}{(m+n-\frac{r^2}{l^2})}~.
\end{aligned}
\ee
The Lagrangian density (\ref{d3LDensity}) can now be written as:
\be
\begin{aligned}
{\cal L} &=  l^2 \, r \, \partial_\theta r ( \sin^2\theta \, \partial_{\phi_1} \xi - \cos^2 \theta \, \partial_{\phi_2} \xi) + r^2 \, \left( (l^2 + r^2) ( \partial_{\phi_1} \xi + \partial_{\phi_2} \xi) + r^2 \partial_{\phi_0} \xi \right) \, \cos\theta \, \sin\theta \\
&\hspace{1cm}
+   r^4 \, \cos\theta \, \sin\theta ~,
\end{aligned}
\ee
where the first line is from the DBI part and the second from the WZ part of the action. Now substituting $\partial_{\phi_1} \xi = m$, $\partial_{\phi_2} \xi = n$ and $\partial_{\phi_0} \xi = -(m+n+1)$ that follow from the monomial solutions, we have the on shell action to be
\be
\begin{aligned}
{\cal L} &=l^2 \, r \, \partial_\theta r (m \, \sin^2\theta \, - n \, \cos^2 \theta ) +(m+n) \,  l^2 \, r^2  \, \cos\theta \, \sin\theta \cr 
&= \partial_\theta \Big[ \frac{l^2}{2} (m \, \sin^2\theta \, - n \, \cos^2 \theta ) \, r^2 \Big] = \partial_\theta \Big[  \frac{1}{2} r \, \Pi_r^\theta \Big]~.
\end{aligned}
\ee
Therefore the modified Lagrangian density ${\cal L} - \partial_\theta \Big[ \frac{1}{2} r \, \Pi_r^\theta \Big]$ vanishes on-shell for any $(m,n)$ locally. Equivalently, we can add a boundary term 
\be
{\cal L}^{(1)}_{bdy} = \frac{1}{2}\, r\, \Pi_r^{\theta} ~,
\ee
near the boundary, which for the cases we shall consider, corresponds to $\theta=0$. 

This completes our analysis of the variational problem for monomial solutions. We note in passing that the boundary term we have obtained is similar to the one obtained in \cite{Drukker:2005kx} in the context of Wilson loops and their holographic realization in terms of probe D3 branes. 

Just as in the boundary theory, one can check that the classical phase space variables satisfy constraints on-shell, which we list below:
\be
\begin{aligned}
{\cal C}_1:&\quad \cos^2\theta\, \Pi^{\phi_1}_{\xi}+\sin^2\theta \, \Pi^{\phi_2}_{\xi} -\Pi^{\tau}_{\xi} -l^2r^2\sin\theta\cos\theta = 0~,\\
{\cal C}_2:& \quad r\, \Pi^\theta_r\left(1+ \frac{l^2}{r^2}-\frac{1}{r^4}(\cot\theta\, \Pi^{\phi_1}_{\xi} +\tan\theta \, \Pi^{\phi_2}_{\xi} ) \right)+\sin\theta\cos\theta(\Pi^{\phi_1}_{\xi} -\Pi^{\phi_2}_{\xi} )  =0~.
\end{aligned}
\label{bulkconstraints}
\ee
These constraints are a non-trivial consequence of the D3-branes being supersymmetric and one can verify them easily using the expressions in \eqref{Pisinbulk}. As in the boundary theory these will prove useful in regularizing the energies. 

\subsection{Renormalized Energies}

We now show that, for a subset of cases, it is possible to perform a holographic renormalization of the energy of the probe D3 branes. In particular we show that by adding further boundary terms that are proportional to the phase space constraints (so that the vanishing result for the on-shell action is unaffected), it is possible to regulate the energies for a number of cases. The analysis will, to some extent, be parallel to the one that we carried out in the Yang-Mills theory in Section \ref{YMregular}. We work explicitly with the three cases listed previously. 

\subsubsection{The static case $m+n+1=0$}

We begin with the static solution with $m+n+1=0$ and we set $m>0$ and $n<0$. For this time-independent case, as we have already mentioned, the energy coincides with the on-shell action and we obtain zero energy. This is perfectly consistent with the results of the boundary theory.

\subsubsection{The probes of type $(0,n)$}

As a first non-trivial case, we consider the defects defined by the integers $(0,n)$ with $n<0$. For this case we find that the integral of the energy density is naively divergent. We now recall two important points: firstly, in the limit that $\frac{l}{R_0}\rightarrow 0$, the probe brane profie $r(\theta)$ exactly coincides with the boundary profile $|Z|$ for the particular solution under consideration. Secondly, for this case of the $(0,n)$ BPS string, the boundary action is particularly simple (see equation \eqref{0nbdyaction}). Given these, we propose the following boundary term for this case: 
\be
\label{bdyterm0n}
L_{bdy}^{(2)} = \frac{1}{2}\tan\theta\, {\cal C}_1 ~,
\ee
where ${\cal C}_1$ is the constraint defined in \eqref{bulkconstraints}. 
The vanishing of the action on-shell is unaffected by the addition of this term as the phase space combination in ${\cal C}_1$ vanishes on-shell. Furthermore, by explicit calculation we have checked that the additional piece exactly cancels the power law divergences as $\theta\rightarrow 0$. The additional term, on account of the presence of $\Pi^{\tau}_{\xi}$ in the constraint, does modify the energy and, after including the overall factors of the tension of the D3-brane $T_{D3}= \frac{N}{2\pi^2 l^4}$ and the $4\pi^2$ coming from the angular integration, we find that for the $(0,n)$ case, the energy takes the following form:
\be
 \frac{l\, E_{0,n}}{4\pi^2} = \frac{N}{2\pi^2}\left( (n+1) \frac{R_0^2}{2l^2} +\frac{1}{2}(n-1)(n+1)^2 + O(\frac{l}{R_0})\right)
 \ee
We now use the map between the parameters of the probe brane and the gauge theory results that was discussed for the half-BPS case in \eqref{mapofparam}. In this case we have 
\be
\label{Rvsr}
R_0 = \frac{2\pi}{\sqrt{\lambda}}\, l\, r_0~,
\ee
and with this map we can rewrite the energy in variables suitable for comparison with the gauge theory side: 
\be
 \frac{l\, E_{0,n}}{4\pi^2} = \frac{1}{g_{YM}^2} \left( (n+1) r_0^2 +\frac{\lambda}{4\pi^2}(n-1)(n+1)^2 + \ldots \right)
   \ee
We find that the leading term exactly matches what we obtained for the energy in \eqref{E0n} and in addition, obtain the leading $O(\lambda)$ quantum correction:

\subsubsection{The probes of type $(-n,n)$}

As the last example, we consider the $(-n,n)$ case for which the exact solution for $r(\theta)$ is given in \eqref{rformnzero}:
 \be
 r(\theta) = \sqrt{R_0^2 \, \cot^{2|n|} \theta - l^2} ~.
 \ee
For negative $n$ the range of $\theta$ is $(0, \theta_0^{(n)})$ where $\cot^{2|n|} \theta_0^{(n)} = l^2/R_0^2$. The energy contribution of the bulk after integration over $\theta$ is (upto the factor $4\pi^2 \, T_{D3}$)
\begin{multline}
l\, E(-n,n,\theta)= \frac{1}{2} \cos^2\theta_0^{(n)} - \frac{R_0^2}{2l^2 (1-n)} \cos^{2-2n}\theta_0^{(n)} \, F(1-n,-n, 2-n, \cos^2\theta_0^{(n)}) \\
 - \frac{1}{2} \cos^2\theta + \frac{R_0^2}{2l^2 (1-n)} \cos^{2-2n}\theta \, F(1-n,-n, 2-n, \cos^2\theta) \Big) ~.
\end{multline}
with $0< \theta < \theta_0$. Here we once again denote the hypergeometric function ${}_2F_{1}$ as simply $F$. The contribution of the first boundary term $-\frac{1}{2} r \, \Pi_r^\theta$ at $\theta$ is
\be
l\, E^{(1)}_{bdy} = \frac{ \, n \cos^2\theta \, (R_0^2 - l^2 \cot^{2n}\theta)^2}{2 \, \left(\cos^2\theta \, (R_0^2 - l^2 \cot^{2n}\theta)^2 + l^2 n^2 R_0^2 \cot^{2n}\theta \, \csc^2\theta \right)}~.
\ee
In the limit that $\theta \rightarrow 0$ (which is the location of the boundary), this term is completely regular for $n<0$ and has the following limiting values: 
\be
l\, E^{(1)}_{bdy} = 
\begin{cases}
\phantom{- -} 0 & \text{for $-1<n<0$.} \\
- \frac{ R_0^2}{2 (l^2 + R_0^2)} & \text{for $n=-1$.}\\
\phantom{- -} \frac{n }{2}  & \text{for $n < -1$.}
\end{cases}
\ee
As in the corresponding boundary problem, one needs to add an additional contribution in order to cancel the power law divergences in this case. We propose the following boundary term 
\be
\label{mnzerobdyterm}
{\cal L}^{(2)}_{bdy} = \frac{1}{l^3}f(n,\theta) \, {\cal C}_1~,
\ee
where ${\cal C}_1$ is the constraint in \eqref{bulkconstraints} and the function $f(n,\theta)$ takes the following form, inspired largely by the corresponding term in the boundary theory (compare with equation \eqref{fmboundary}, setting $m=-n$):
\be
f(n,\theta) = \frac{1}{2(1+n)} \tan\theta \, F(1,n, 2+n, - \tan^2\theta)~.
\ee
This term leads to an additional contribution to the energy and taking into account the bulk and boundary contribution, we finally obtain 
\begin{multline}
\label{Eminusnbulk}
l\, E_{-n,n} = \frac{n-1}{2} + \frac{R_0^2 }{2l^2} \Gamma(1-n) \Gamma(1+n) \cr
 + \frac{1}{2} \cos^2\theta_0^{(n)} - \frac{R_0^2}{2l^2 (1-n)} \cos^{2-2n}\theta_0^{(n)} \, F(1-n,-n, 2-n, \cos^2\theta_0^{(n)})~.
\end{multline}
In order to make contact with the results of the boundary theory, we take the limit in which $\frac{l}{R_0}\rightarrow 0$. In this limit, as before, the profile of the D3 brane $r(\theta)$ coincides with boundary profile $|Z|$ of the boundary theory.  This in turn corresponds to $\theta_0^{(n)}\rightarrow \frac{\pi}{2}$ which sets the terms in the second line of \eqref{Eminusnbulk} to zero. 

As we saw in the boundary theory regarding the energies of BPS strings for the $(-n,n)$ case, the result is divergent if $n$ takes negative integer values, which suggests that one would need to add additional terms to deal with such divergences. However for fractional values of $n$, we obtain finite values. By once again taking into account the map \eqref{Rvsr} between the bulk and boundary parameters, and restoring the factor of $4\pi^2T_{D3}$, we find that:
\begin{align}
\left(\frac{l\, E_{-n,n}}{4\pi^2}\right)_{n=-\frac{(2p+1)}{2}} &=(-1)^{p} \frac{\pi\, (2p+1)}{2g_{YM}^2}\, r_0^2+\frac{N}{4\pi^2}(n-1)\\
&=  \frac{1}{g_{YM}^2} \left((-1)^{p} \frac{\pi\, (2p+1)}{2}\, r_0^2 + \frac{\lambda}{4\pi^2}(n-1) \right)~.
\end{align}
In this case we started with the {\it exact} solution for the profile $r(\theta)$. We find a perfect match at the leading order with the gauge theory answer in \eqref{E-nn} and in addition we obtain the first quantum correction to the energy of the BPS string. 

\subsubsection{The general $(m,n)$ probe} 
We now consider the general case with $(m>0,n<0)$. In general we do not have a closed form expression for $r(\theta)$. Instead we can solve \eqref{rdef} by a series solution:
\bea
r(\theta) =\widehat r(\theta) - \frac{l^2}{\widehat r (\theta)} \sum_{q=1}^\infty \frac{1}{q! \, 2^q \, (2q-1)} \left( \frac{l^2}{\widehat r^2(\theta)} \right)^{q-1} \, \prod_{r=1}^{q} \Big( (2q-1)(m+n)+ (2r-1) \Big) ~,
\eea
where $\widehat r(\theta) = R_0 \, \cos^m\theta \, \sin^n \theta$. This series is obtained by expanding the differential equation for $r(\theta)$ for $r(\theta) >> l$ and the result is a useful one when $\frac{R_0}{l} >>1$ and generic values of $\theta$ (away from zeros and singularities of $\widehat r(\theta)$). 

The rest of the analysis in this section closely parallels our discussion of the strings with $m+n=0$, except that, instead of the exact solution, we make use of the series expansion. Using the series expansion explicitly and performing the worldvolume integrals, we find that the contribution of the bulk and the first boundary term to the energy, after integrating up to $\theta$ is (up to the factor of $4\pi^2 T_{D3}$):
\begin{multline}
\label{bulkenergymn}
l\, E(m,n,\theta)= \frac{R_0^2}{2l^2(1+m)} (m+n+1)^2 \cos^{2m+2} \theta \, F(1+m, -n, 2+m, \cos^2\theta) \cr + \frac{1}{2} (m+n+1)^2 (m+n-1) + {\cal O}\left(\frac{l^6}{R_0^2}\right)~,
\end{multline}
where $0 < \theta < \pi/2$ (except when $m+n=0$ as discussed already). We now add the boundary term 
\bea
{\cal L}_{bdy}^{(2)} = \frac{1}{l^3} f(m,n,\theta) \, {\cal C}_1~,
\eea
with 
\be
\label{mnbdyterm}
f(m,n,\theta) = \frac{m+n+1}{2(1+n)} \tan\theta \, F(1,-m, 2+n, - \tan^2\theta)~.
\ee
We note that when $m+n=0$, this reduces to the boundary term in \eqref{mnzerobdyterm}. Similarly, for $m=0$, the hypergeometric term reduces to unity and the boundary term reduces to the one in \eqref{bdyterm0n}. Thus, what we have in \eqref{mnbdyterm} is the general boundary term that is applicable to all the monomial cases. 
It contributes the following term to the energy (in units of $1/l$):
\bea
\frac{R_0^2 (m+n+1)^2}{2l^2(1+n)} \tan\theta \, F(1,-m, 2+n, - \tan^2\theta) \,  \cos^{2m+1}\theta \, \sin^{2n+1} \theta  + {\cal O}\left(\frac{l^6}{R_0^2} \right)~.
\eea
Just as in the field theory calculation of Section \ref{giantdefects} (see for instance equations \eqref{mnsubs}) we replace $(m,n)$ by the appropriate combinations of the fields:
\be
m \, l^2 \rightarrow \frac{\Pi^{\phi_1}_\xi \, r^2 \, \cot\theta}{\Pi^{\phi_1}_\xi \, \cot\theta + \Pi^{\phi_2}_\xi \, \tan\theta -r^2 ( l^2  + r^2) }, ~~ n \, l^2 \rightarrow \frac{\Pi^{\phi_2}_\xi \, r^2 \, \tan\theta}{\Pi^{\phi_1}_\xi \, \cot\theta + \Pi^{\phi_2}_\xi \, \tan\theta -r^2 ( l^2  + r^2) }~.
\ee
One can then check that these boundary terms do remove the singular pieces from (\ref{bulkenergymn}) and renders the energy finite up to ${\cal O}\left(\left(l/R_0\right)^0 \right)$. Restoring the factors of $4\pi^2 T_{D3}$ we finally obtain
\begin{align*}
\frac{l\, E_{m,n}}{4\pi^2} = \frac{N\, R_0^2 }{4\pi^2 l^2} (m+n+1) \frac{\Gamma(1+m) \Gamma(1+n)}{\Gamma(1+m+n)} + \frac{N}{4\pi^2 } (m+n+1)^2 (m+n-1) + {\cal O}\left(\frac{l^2}{R_0^2} \right)~,
\end{align*}
where, as before, the first term has to be analytically continued whenever it does lead to a finite answer. Rewriting in terms of the gauge theory parameters we find that 
\begin{align}
\label{Efinal}
l\, E_{m,n} = \frac{4\pi^2 r_0^2 }{g_{YM}^2} (m+n+1) \frac{\Gamma(1+m) \Gamma(1+n)}{\Gamma(1+m+n)} + \frac{\lambda}{4\pi^2 g_{YM}^2} (m+n+1)^2 (m+n-1) + \ldots
\end{align}
At higher orders in $\frac{l^2}{R_0^2}$ we expect that the singularities, if any, coming from the regions close to the boundary can also be removed using our methods.

Let us note that the holographic renormalization that we carried out for the probe brane provides a justification for the regularisation of the energies that was done on the CFT side in Section \ref{giantdefects}. 
However there are a few important questions that our analysis leaves open. It is fair to say that, while our regularization does indeed cancel all coordinate dependent power law divergences there is a certain ambiguity in the finite part of the charges. We have chosen the boundary terms so that the energies of these states are an analytic continuation in the $(m,n)$ parameters that appear in the monomial defining the BPS string. This can be seen clearly in \eqref{Efinal}. For both $(m,n)$ negative and integer, or for particular cases such as $m+n=0$ with the parameters being integers, the energies calculated using both the gauge theory and probe brane analyses are divergent and it is unclear how to treat these constant divergences that appear for particular negative integer values of $m$ and $n$. Further, in those cases for which the energy $E_{m,n}$ is finite, the interpretation of the energy remains an open problem. Given that the sign of the energy can be either positive or negative it is likely that these can be interpreted as Casimir energies of some effective theory on the BPS string.

\section{Summary and Discussion}
\label{SandD}

We have analyzed classical $\frac{1}{8}$-BPS configurations in the abelian sector of ${\mathcal N}=4$ supersymmetric Yang-Mills theory on ${\mathbb R} \times S^3$ that correspond to time-dependent string like defects -- which we termed wobbling BPS strings. These strings are characterized by singular profiles for one of the complex scalar fields of the gauge theory at the location of the defect. 
Below we conclude with a brief summary of our observations regarding these wobbling BPS strings and a discussion of some open questions and possible future work. 

The $\frac{1}{8}$-BPS wobbling strings on ${\mathbb R} \times S^3$ are related by Wick rotation and Weyl transformation to $\frac{1}{8}$-BPS defects associated with surface operator in ${\mathbb R}^4$. Hence our BPS strings should be understood as the states corresponding to the Gukov-Witten type surface operators and their lower supersymmetric generalizations dictated by the state-operator correspondence of the CFT. One of the main goals of this work and that we have been able to achieve, is a characterization of the general BPS string solution that preserves four supercharges. The location of such a BPS string at any instant of time is obtained as the intersection of zeros of holomorphic functions in ${\mathbb C}^2$ with $S^3\subset {\mathbb C}^2$. The time evolution of the string is also simply obtained from our analysis. Moreover, the description of the string solution is recovered from the holographic side by analyzing the worldvoume constraints on probe D3-branes in $AdS_5\times S^5$ and studying the limiting behaviour near the boundary of $AdS_5$. 

The holographic duals of our $\frac{1}{8}$-BPS strings preserve precisely the same supersymmetries as the $(S_1, S_2, J)$ giants of \cite{Mandal:2006tk} and the wobbling dual-giants of \cite{Ashok:2008fa}. By the addition of appropriate boundary terms we showed that the abelian solutions that are regular (dual to the dual-giants) as well as the singular string solutions of the CFT can be made to belong to the same variational problem. We then showed that the singularities in the classical expressions of the energy and other charges can be systematically ``renormalized" by including additional boundary terms on the CFT side. The holographic dual of this procedure was carried out for the monomial type D3-brane probes. The analysis of the probe brane theory paralleled that in the Yang-Mills theory and the leading order results for the energy and charges could be matched once the parameters of the solution were appropriately mapped to each other. 

In previous work \cite{Ashok:2010jv}, a generic description of electromagnetic waves on the $\frac{1}{8}$-BPS wobbling dual-giants was provided. Now that the holographic duals of BPS strings of the CFT also belong to the same supersymmetry class as the wobbling dual-giants, it follows that one can also turn on electromagnetic waves on the new D3 probes without breaking supersymmetry. This provides (infinitely many) additional parameters to describe the corresponding wobbling string. It is natural to ask what these additional modes correspond to on the CFT side. To answer this question (at least in part) let us note that the  duals of the relevant D3-branes in the bulk belong to the abelian sector of the CFT. Since in the abelian sector the scalars and the gauge fields decouple one can turn on the pure-glue solutions discussed in Appendix \ref{SO6class} without affecting the profiles of the scalars (or vice versa). It should be possible to set-up a detailed correspondence between the EM waves on the D3-branes in the bulk theory and the pure-glue dressing of the scalar solutions of the boundary theory. Furthermore, through the operator-state correspondence, this predicts as many additional parameters to characterise the analogous $\frac{1}{8}$-BPS surface operators in the Euclidean theory and it will be interesting to explore this in more detail.

We considered BPS string defects supported mainly by a single scalar (and gauge fields). Therefore these can be embedded into many other CFTs with less supersymmetry. For instance the holographic dual D3-branes we considered can be embedded into less supersymmetric spacetimes such as $AdS_5 \times {\cal M}^5$, for some suitable Sasaki-Einstein 5-manifold ${\cal M}^5$ in a straightforward manner (see for instance \cite{Koh:2009cj}). Therefore all our computations of holographic renormalization etc. can be adapted easily to the corresponding bulk as well as boundary theories. 

While we obtained a general characterization of the wobbling BPS strings, it would be important to have a more detailed understanding of the space of solutions to these equations as it could have interesting consequences for the physics of these defects. For instance, it is known in the mathematics literature that the solutions to the intersection of the zeros of a holomorphic function with $S^3$ include quasi-positive algebraic links \cite{Milnor, Hayden}. This is especially interesting in the context of work relating four dimensional gauge theory and knot theory in the Euclidean context.  By studying the gauge theory on a four dimensional half space \cite{Witten:2011zz, Henningson:2011qk, Henningson:2011ak} it has been shown that solutions of the generalized Bogomolny equations \cite{Kapustin:2006pk} that correspond to codimension two defects with singular boundary conditions along a knot can be used to study topological invariants associated to the knot such as the Jones polynomial and play an important role the programme of categorification \cite{Gukov:2007ck, Witten:2011zz, Gaiotto:2011nm, Mazzeo:2017qwz}. One would hope that the Hamiltonian analysis of the  supersymmetric sector that includes these BPS strings in the physical ${\mathcal N}=4$ theory might also prove useful in these efforts. Apart from the classification problem, it would also be worthwhile to explore particular solutions that might have important physical applications. For instance one could check if the solution space to these equations for BPS strings include configurations that self intersect and that could be interpreted as junctions or networks of defects \cite{Chun:2015gda}. 

We have studied only abelian solutions throughout this work and this includes the pure glue defects that are discussed briefly in Appendix \ref{SO6class}. However it is important point to keep in mind that the $\frac{1}{8}$-BPS equations that were derived in \eqref{finalBPS} are fully non-abelian. Given that the same equations allow for both local operators as well as string like solutions, and given that they are part of the same supersymmetric sector, it is natural to look for possibly non-abelian solutions that interpolate between the two classes of solutions. Analogous questions in the bulk would involve finding a probe D3 brane that interpolates between a giant or dual-giant graviton and a noncompact probe dual to a wobbling string solution.

We end with a few remarks about the relevance of these wobbling string solutions to the quantum Yang-Mills theory. 
We have carried out a (semi-) classical analysis of these singular string solutions on the CFT side. It is natural to wonder whether one can construct the BPS strings as some type of boundary states using the perturbative BPS states of the gauge theory. Lastly it is reasonable to expect a low energy description of the wobbling strings in terms of some suitable degrees of freedom living on the defect. It will be interesting to seek such a description.

\vskip 1cm
\noindent {\large {\bf Acknowledgments}}
\vskip 0.2cm
We would like to thank Marco Bill\`o, Marialuisa Frau, Dileep Jatkar, Renjan John, Alok Laddha, Alberto Lerda, Madhusudhan Raman and Ranadeep Roy for helpful discussions. We thank Rohan Poojary for collaboration on a  previous unpublished work. We are especially grateful to Tudor Dimofte and Parameswaran Sankaran for helpful correspondence. 
\vskip 1cm

 \begin{appendix}

 \section{Pure Glue Defects}
\label{SO6class}

We now briefly consider the class of classical solutions in which the scalar fields are set to zero and only gauge fields are turned on. The BPS equations take the simplified form:
\be
\begin{aligned}
F_{12} &= F_{03} = 0~, \qquad F_{01}+F_{31}=0~, \qquad F_{02}+F_{32} = 0~.
\end{aligned}
\label{maxwell-bpseqns}
\ee
This implies that the independent components of the field strengths are $F_{01}$ and $F_{02}$. Although these were obtained as part of the $\frac{1}{8}$-BPS projections, it is straightforward to check that these equations actually preserve eight supercharges and are $\frac{1}{4}$-BPS configurations. They preserve the supersymmetries that survive the following projections:
\be
(1+\Gamma^{03})\eta^{(-)}_A = 0 \quad\text{for}\quad A=1,2,3,4~.
\ee

Our goal is to solve for the field strengths in this gauge sector and check if there are singular solutions, analogous to the ones we have found in the scalar sector. From the discussion of the equations of motion and Bianchi identities in Section \ref{EoMBI}, we recall that there are two additional differential constraints that arise, which take the form:
\be
(D_0+D_3+i)(F_{01} - i F_{02}) = 0~, \qquad \left( D_1 + i  D_2 \right) \left( F_{01} - i F_{02} \right) = 0~.
\ee
In the abelian case, the gauge and local Lorentz covariant derivative $D_a$ coincides with the vector fields $E_a$ defined in \eqref{hopfvectorfields} and we observe that the equations satisfied by the complex combination $F_{01}-i F_{02}$ is identical to those satisfied by the scalar field $Z$ in \eqref{Z1eqns}. Thus, the most general solution is given by
\be
\label{maxwell-finalsolutions}
F_{01}-i F_{02}= f (\zeta_1, \zeta_2) ~,
%\sum_{mn} c_{m,n} \, \left(\cos\tfrac{\vartheta}{2} \, e^{i \, \frac{\tilde \psi + \phi}{2}} \right)^m \left(\sin \tfrac{\vartheta}{2} \, e^{i \, \frac{\tilde \psi - \phi}{2}} \right)^n~.
\ee
where we have defined the $\zeta_i$ (as in \eqref{zetadefn}) to be 
\be
\zeta_1 = \cos\theta\, e^{i \,(\phi_1 - \tau)}\qquad \zeta_2 = \sin\theta \, e^{i \,(\phi_2 - \tau)} ~.
\ee
For a monomial solution of the form $f = c_{m,n}\zeta_1^m\, \zeta_2^n$, depending on the values of $(m,n)$ the solution is regular or singular similar to those in (\ref{solution1}). We have time-independent monomial solutions for $m+n=0$. The conserved charges of these monomial solutions can be derived from the stress tensor and these are given by (see Appendix \ref{Tmunu} for details):
\be
\begin{aligned}
E &= \frac{2\pi^2}{g_{YM}^2} |c_{mn}|^2 (m+n)  \frac{\Gamma(m) \Gamma(n)}{\Gamma(m+n+1)}, \cr 
S_1 &= - \frac{2\pi^2}{g_{YM}^2} |c_{mn}|^2 \, m \frac{\Gamma(m) \Gamma(n)}{\Gamma(m+n+1)}\cr
S_2 &= - \frac{2\pi^2}{g_{YM}^2} |c_{mn}|^2 \, n \frac{\Gamma(m) \Gamma(n)}{\Gamma(m+n+1)}~.
\end{aligned}
\label{ES1S2gauge}
\ee
for positive $(m,n)$. They clearly satisfy the linear relation $E + S_1 + S_2 =0$. These solutions preserve eight supercharges of the ${\cal N}=4$ SYM and share four supercharges with those of (\ref{solution1}). The equations and solutions in this sector for positive $(m,n)$ have been analyzed in detail in \cite{Grant:2008sk, Yokoyama:2014qwa}. 

For non-positive $m$ or $n$ one again has singular solutions -- with the singularity in the field configurations generically extending along a two-dimensional subspace -- analogous to the defects supported by the scalar field $Z$ and so we refer to these as pure-glue defects. Unlike the scalar defects here the singular and non-singular solutions already belong to the same variational problem since the BPS conditions (\ref{maxwell-bpseqns}) ensure that the Lagrangian density vanishes for all the solutions we have. It will be interesting to see if the charges $E, S_1, S_2$ can also be regularized for the singular defects as well. 

We conclude this discussion by using the Wick rotation and Weyl transformation to express the solutions in  Euclidean space $\mathbb{R}^4$, with coordinates $(z_1, z_2)$. The general solution in \eqref{maxwell-finalsolutions} maps to the following non-vanishing components in Euclidean space:
\be
\label{Ff}
\begin{aligned}
F_{12} &= \frac{1}{z_1z_2} f(z_1,z_2)~, \qquad 
F_{1\overline 1} = -F_{2\overline 2}  = \frac{1}{r^2}\,  \overline{f}\left(\frac{\bar z_1}{r^2},\frac{\bar z_2}{r^2}\right) \\
 F_{1\overline 2} &= -\frac{\bar z_1}{\bar z_2}\, \frac{1}{r^2}\, \overline{f}\left(\frac{\bar z_1}{r^2},\frac{\bar z_2}{r^2}\right)~, \qquad  
 F_{\overline 1 2} = -\frac{\bar z_2}{\bar z_1}\, \frac{1}{r^2}\, \overline{f}\left(\frac{\bar z_1}{r^2}\, \frac{\bar z_2}{r^2}\right)~.
\end{aligned}
\ee
The different scaling of $z_i$ and $\bar z_i$ are a simple consequence of the interpretation of the radial direction in Euclidean space with the Wick rotated time coordinate $e^{\tau_E}$. The field strength components in turn can be shown to satisfy the following BPS equations in Euclidean space \cite{Grant:2008sk}:
 \be
 \begin{aligned}
 & F_{\overline 1\overline 2} = 0 ~, \quad F_{1\overline1} + F_{2\overline 2} = 0~, \\
\text{and}\quad &\sum_{j=1}^2 \bar z_j F_{i\overline j} = 0  \quad\text{for $i = 1, 2$}~.   
 \end{aligned}
 \ee
It would be an interesting problem to better understand the solutions \eqref{Ff} for functions that lead to codimension two defects (as for the scalar sector), and interpret them as pure glue surface defects in $\mathbb{R}^4$.
 
 \section{Stress Tensor and Charges}
\label{Tmunu}

In this section we obtain the stress tensor for the scalar and gauge sector of ${\mathcal N}=4$ Yang-Mills theory on $\mathbb{R}\times S^3$ and obtain the charges associated to the spacetime symmetries by integrating the current densities over the $S^3$ factor.
 
 \subsection{The scalar sector}
 
 The Lagrangian density for a conformally coupled massless complex scalar field on a four dimensional manifold with metric $g_{\mu\nu}$ is given as follows:
\be
%\label{cft-lagrangian}
{\cal L} = - \frac{1}{g_{YM}^2} \sqrt{-g} \left[ g^{\mu\nu} \partial_\mu Z \partial_\nu \bar Z + \frac{1}{6}\, {\mathcal R}\, \bar Z \, Z \right]~,
\ee
where ${\mathcal R}$ is the Ricci scalar of the manifold. The standard result for the stress tensor is obtained by varying the metric and computing the resulting variation of the action. The resulting symmetric stress tensor is given by
\begin{multline}
\label{CCT}
T_{\mu\nu} =-\frac{1}{g_{YM}^2}(\pa_\mu Z\pa_\nu \bar Z - \pa_\nu Z\pa_\mu\bar Z)+\frac{1}{g_{YM}^2}g_{\mu\nu}g^{\kappa\lambda}\pa_\kappa Z\pa_\lambda \bar Z - \frac{1}{3g_{YM}^2}(g_{\mu\nu}\Box - \nabla_\mu\nabla_\nu)(Z\bar Z) \\
- \frac{1}{3g_{YM}^2}(R_{\mu\nu}-\frac{1}{2}g_{\mu\nu}{\mathcal R}) (Z\bar Z)~.
\end{multline}
We choose the metric on $ {\mathbb R}\times S^3$ to be
\be
ds^2 = - d\tau^2 + (d\theta^2 + \cos^2\theta \, d\phi_1^2 + \sin^2\theta \, d\phi_2^2)~,
\ee
and substitute in the expression in \eqref{CCT} to find the resulting traceless symmetric stress tensor. 

The charges that we are after are given by the following spatial integrals:
\be
\begin{aligned}
 E = \int_{S^3}\  \sqrt{g}\ T^\tau_\tau~,\quad S_1 = \int_{S^3}  \sqrt{g}\ T^\tau_{\phi_1}~, \quad S_2 = \int_{S^3}\ \sqrt{g}\ T^\tau_{\phi_2}~.
\end{aligned}
 \ee
We then substitute  the solutions of interest, namely
$$
Z = r_0\, (\cos\theta e^{i\phi_1})^m\, (\sin\theta e^{i\phi_2})^n~,
$$
into the expression for the charges. The integrals are straightforward to perform and we obtain
where $B(a,b)$ is the Euler beta function. The other charges for $m,n \ge 0$ are computed from the stress tensor (see Appendix \ref{Tmunu}) and we obtain
\be
\begin{aligned}
E(r_0, m,n) &= \frac{4\pi^2 \, r_0^2}{g_{YM}^2} \, {(m+n+1)} \frac{\Gamma(1+m)\Gamma(1+n)}{\Gamma(m+n+1)}\\
% \, \cr &=  2\pi^2 \, r_0^2 \, (m+n+1)^2 \, B(m+1,n+1)~,\\
%J(r_0,m,n) &= 2\pi^2 \, r_0^2 \, (m+n+1) \, B(m+1,n+1)~, \\
S_1(r_0,m,n) &= -\frac{4\pi^2 \, r_0^2}{g_{YM}^2} \,  m \,\frac{\Gamma(1+m)\Gamma(1+n)}{\Gamma(m+n+1)}~,\\
S_2(r_0,m,n) &= -\frac{4\pi^2 \, r_0^2}{g_{YM}^2} \,  n \, \frac{\Gamma(1+m)\Gamma(1+n)}{\Gamma(m+n+1)}~.
\end{aligned}
\label{ES1S2appendix}
\ee
These are the results as stated in \eqref{ES1S2}.
 
 \subsection{The gauge sector}

We have only worked in the abelian sector and the Maxwell Lagrangian density is given by
\be
{\mathcal L} = -\frac{1}{4g_{YM}^2}\sqrt{g}\, F^{\mu\nu}F_{\mu\nu}~. 
\ee
The $\frac{1}{4}$-BPS solutions are written most simply in the frame basis and this is what is written in the main text. Translating these into the coordinate basis we find that 
\be
\begin{aligned}
F_{\tau\phi_1} &=- F_{\tau\phi_2}  \quad F_{\tau\theta} =\csc^2\theta\, F_{\theta\phi_2} \\
F_{\theta\phi_1} &= \cot^2\theta F_{\theta\phi_2}\quad F_{\phi_1\phi_2} = - F_{\tau\phi_2}~.
\end{aligned}
\ee
Thus, there are two independent  components of the gauge field. These are constrained by the equations of motion and Bianchi identities. As we show in the main body of the paper, only four of these impose any new conditions. Two of these can be solved by parametrizing the independent components as follows:  
\be
F_{\tau\phi_2} = f_\tau(\theta, \phi_1-\tau,\phi_2-\tau)\qquad F_{\theta\phi_2} = \tan\theta f_\theta(\theta, \phi_1-\tau,\phi_2-\tau)
\ee
 The remaining two conditions in turn can be solved in terms of linear combinations of the following basic solutions:
 \be
 \begin{aligned}
 f_\tau(\theta, \phi_1,\phi_2) + i   f_\theta(\theta, \phi_1,\phi_2) &= c_{m,n}\, (\cos\theta)^m(\sin\theta)^n \, e^{i m \phi_1+i n \phi_2}~.
% f_\theta(\theta, \phi_1,\phi_2) &= (\cos\theta)^m(\sin\theta)^n(\lambda_{2,m,n} \cos(m\phi_1+n\phi_2)+\lambda_{1,m,n}\sin(m\phi_1+n\phi_2) 
 \end{aligned}
 \ee
These are the components of the solutions to the BPS equations in the coordinate basis. The stress tensor for the Maxwell theory is given by the expression
\be
T^{\mu}_{\nu} = \frac{1}{g_{YM}^2}\left(F^{\mu\lambda}F_{\lambda\nu} - \frac{1}{4}g_{\mu\nu}F^{\lambda\rho}F_{\lambda\rho}\right)~.
\ee
Substituting the solution into the stress tensor we find that the second term proportional to the metric vanishes. Integrating the appropriate components over the $S^3$, we find the following charges:
\be
\begin{aligned}
 E &= \int_{S^3\ }\sqrt{g}\ T^{\tau}_{\tau} =  \frac{2\pi^2}{g_{YM}^2} |c_{mn}|^2 (m+n)  \frac{\Gamma(m) \Gamma(n)}{\Gamma(m+n+1)}~, \cr 
S_1 &=  \int_{S^3\ }\sqrt{g}\ T^{\tau}_{\phi_1} =- \frac{2\pi^2}{g_{YM}^2} |c_{mn}|^2 \, m \frac{\Gamma(m) \Gamma(n)}{\Gamma(m+n+1)}~,\cr
S_2 &=  \int_{S^3\ }\sqrt{g}\ T^{\tau}_{\phi_2} =- \frac{2\pi^2}{g_{YM}^2} |c_{mn}|^2 \, n \frac{\Gamma(m) \Gamma(n)}{\Gamma(m+n+1)}~.
\end{aligned}
\ee
These are the charges we have quoted in \eqref{ES1S2gauge}.
 
\end{appendix}
\endgroup

\providecommand{\href}[2]{#2}\begingroup\raggedright\endgroup

\end{document}